\documentclass[10pt,final,letter,twocolumn,twoside,journal,romanappendices]{IEEEtran}
\usepackage{cite}
\usepackage{epsfig}
\usepackage{amsthm}
\usepackage{amsmath,amssymb,amsfonts,amstext,amsbsy,amsopn,dsfont}
\usepackage{cases}
\usepackage{bigints}
\usepackage{sublabel}
\usepackage{xcolor}
\usepackage{array}
\usepackage{cuted}

\newtheorem{theorem}{\textbf{Theorem}}

\newtheorem{lemma}{\textbf{Lemma}}

\newcommand{\defn}{\triangleq}
\newcommand{\dif}{\textmd{d}}

\begin{document}

\title{Ultra-Reliable and Low-Latency Communications Using Proactive Multi-cell Association}

\author{Chun-Hung Liu, \IEEEmembership{Senior Member, IEEE}, Di-Chun Liang, Kwang-Cheng Chen, \IEEEmembership{Fellow, IEEE}, and Rung-Hung Gau, \IEEEmembership{Senior Member, IEEE}  
\thanks{C.-H. Liu is with the Department of Electrical and Computer Engineering, Mississippi State University, Starkville, MS 39762, USA. (e-mail: chliu@ece.msstate.edu)}
\thanks{K.-C. Chen is with the Department of Electrical Engineering, University of South Florida, Tampa, FL 33620, USA. (e-mail: kwangcheng@usf.edu)}
\thanks{D.-C. Liang and R.-H. Gau are with the Institute of Communications Engineering and Department of Electrical and Computer Engineering, National Chiao Tung University, Hsinchu 30010, Taiwan. (e-mail: ldc30108@gmail.com; runghunggau@g2.nctu.edu.tw)}}


\maketitle

\begin{abstract}
Attaining reliable communications traditionally relies on a closed-loop methodology but inevitably incurs a good amount of networking latency thanks to complicated feedback mechanism and signaling storm. Such a closed-loop methodology thus shackles the current cellular network with a tradeoff between high reliability and low latency. To completely avoid the latency induced by closed-loop communication, this paper aims to study how to jointly employ open-loop communication and multi-cell association in a heterogeneous network (HetNet) so as to achieve ultra-reliable and low-latency communications. We first introduce how mobile users in a HetNet adopt the proposed proactive multi-cell association (PMCA) scheme to form their virtual cell that consists of multiple access points (APs) and then analyze the communication reliability and latency performances. We show that the communication reliability can be significantly improved by the PMCA scheme and maximized by optimizing the densities of the users and the APs. The analyses of the uplink and downlink delays are also accomplished, which show that extremely low latency can be fulfilled in the virtual cell of a single user if the PMCA scheme is adopted and the radio resources of each AP are appropriately allocated.  
\end{abstract}


\section{Introduction}\label{Sec:Introduction}
\IEEEPARstart{T}{he} international telecommunication union has identified  ultra-reliable low-latency communication  (URLLC), machine-type communication (mMTC) and enhanced mobile broadband (eMBB) as the three pillar services in the fifth generation (5G) mobile communication that aims to provide good connectivity for many various communication applications \cite{ITU-R2083,GPHSGB18,KCCTZ19}. Among these three services, URLLC remains the most challenging technology due to the need of completely new system design in order to achieve the extremely high system reliability and low latency in 5G cellular systems. Existing mobile communication systems, such as long-term evolution (LTE) systems and its predecessors, were prominently designed to achieve the goal of high throughput in mobile communications, yet they can also achieve highly reliable communications in the physical layer at the expense of complicated \textit{closed-loop} protocol stack to inevitably result in large networking latency of tens to hundreds of milliseconds. This indicates that there exists a tradeoff between high reliability and low latency in system network architecture and subsequent communication protocols of mobile communication networks. Such a reliability-latency tradeoff problem intrinsically impedes the existing cellular systems to extend their services in mission-critical communication contexts with ultra high reliability and low latency constraints, such as wireless control and automation in industrial environments, vehicle-to-vehicle communications, and the tactile internet which allows controlling both real and virtual objects with real-time haptic feedback \cite{SIMSEK16,SYLSCHKCC15}. 

The message transmission time for mission-critical applications needs to be on the order of milliseconds (ms) because the human reaction time is on the order of tens of milliseconds \cite{ITU-T14} or less toward 1 ms for fully autonomous application scenarios. The end-to-end latency of the LTE systems is usually in the range of $30\sim100$ ms, which cannot be further reduced because the backbone network of the LTE systems typically uses a delivery mechanism which is not optimized for latency-sensitive services. To reduce the end-to-end latency in the cellular systems like LTE, it is necessary to fundamentally change the system architecture relying on the closed-loop communication and backbone links. The latency of the backbone link can be significantly reduced by appropriate communication architecture and implementation of network protocols to construct the dedicated connection for URLLC services \cite{SCHULZ17}. To reduce the latency in the physical layer, transmission overhead needs to be suppressed by streamlining the grant-free transmission mechanism of the physical layer access and allocating resources properly \cite{SYQ18}. Nonetheless, reducing the latency in the communication and backbone links is still insufficient to effectively perform ultra low-latency transmission in the current LTE systems because most of the transmission latency is incurred by the control signaling (e.g., grant and pilot signaling usually takes $0.3\sim0.4$ ms per scheduling). Accordingly, the most important and essential means that enables URLLC in 5G heterogeneous cellular networks (HetNets) toward the target latency of one millisecond is to disruptively redesign the transmission protocols in the physical layer of HetNets \cite{SYLSCHKCC15, SCHKCC15}.

\subsection{Motivation and Prior Related Work}\label{SubSec:PriorWorkMotivation}
To effectively reduce latency in 5G HetNets, the essential approach is to adopt \textit{ (feedback-free) open-loop communication}\footnote{The open-loop communication mentioned in this paper is a ``feedback-free" communication technique, i.e., no \textit{immediate} feedback from the receiver side. In light of this, it is also a ``retransmission-free'' communication technique because no retransmission happens between a transmitter and its receiver. Note that the open-loop communication may still need some delayed control signaling in order to successfully perform in a communication system.} so that no retransmission is needed and receivers can save time in performing additional processing and protocol. Open-loop communication has a distinct advantage to significantly reduce control signaling overhead for power control and channel estimation in cellular systems if compared with closed-loop communication. As such, in this paper we focus on how to fulfill URLLC through open-loop communication in a HetNet owing to the fact that extremely reliable open-loop communication is the key to low latency. All the existing URLLC works  in the literature are hardly dedicated to studying the open-loop communication or without retransmission (typically see \cite{AAMKPC18,BLSPDJL18,JNOORSHA19,RDGDGC19,AAGDV18,CSCSCYTQYL19,CSCYO18,MAKFTO19,THJRAO19}). Some of the recent works focus on how to perform URLLC in wireless systems by employing retransmissions, short packet designs and their corresponding estimation algorithms for point-to-point transmission \cite{AAMKPC18,BLSPDJL18,JNOORSHA19,RDGDGC19}. Reference \cite{AAMKPC18}, for example, studied the energy-latency tradeoff problem in URLLC systems with hybrid automatic repeat request (HARQ), whereas reference \cite{BLSPDJL18} proposed an efficient receiver design that is able to exploit useful information in the data transmission period so as to improve the reliability of short packet transmission. 

There are some of the recent works that investigated resource allocation problems in wireless networks under the URLLC constraint. In \cite{AAGDV18}, the authors studied how to minimize the required system bandwidth as well as optimize the resource allocation schemes to maximize URLLC loads, whereas the problem of optimizing resource allocation in the short block length regime for URLLC was investigated in \cite{CSCSCYTQYL19}. Reference \cite{CSCYO18} studied how to jointly optimize uplink and downlink bandwidth configuration and delay components to minimize the total required bandwidth and end-to-end delay. Furthermore, there are few recent works that looked into the URLLC design from the perspective of physical-layer system interfaces and wireless channel characteristics. The recent work in \cite{JJNRLPP18}, for example, adopted coding to seamlessly distribute coded payload and redundancy data across multiple available communication interfaces to offer URLLC without intervention in the baseband/PHY layer design. The problem of how URLLC is affected by wireless channel dynamics and robustness was  thoroughly addressed in \cite{VNSPRGR19}. 

\begin{figure*}[!t]
	\centering
	\includegraphics[height=3.0in,width=\linewidth]{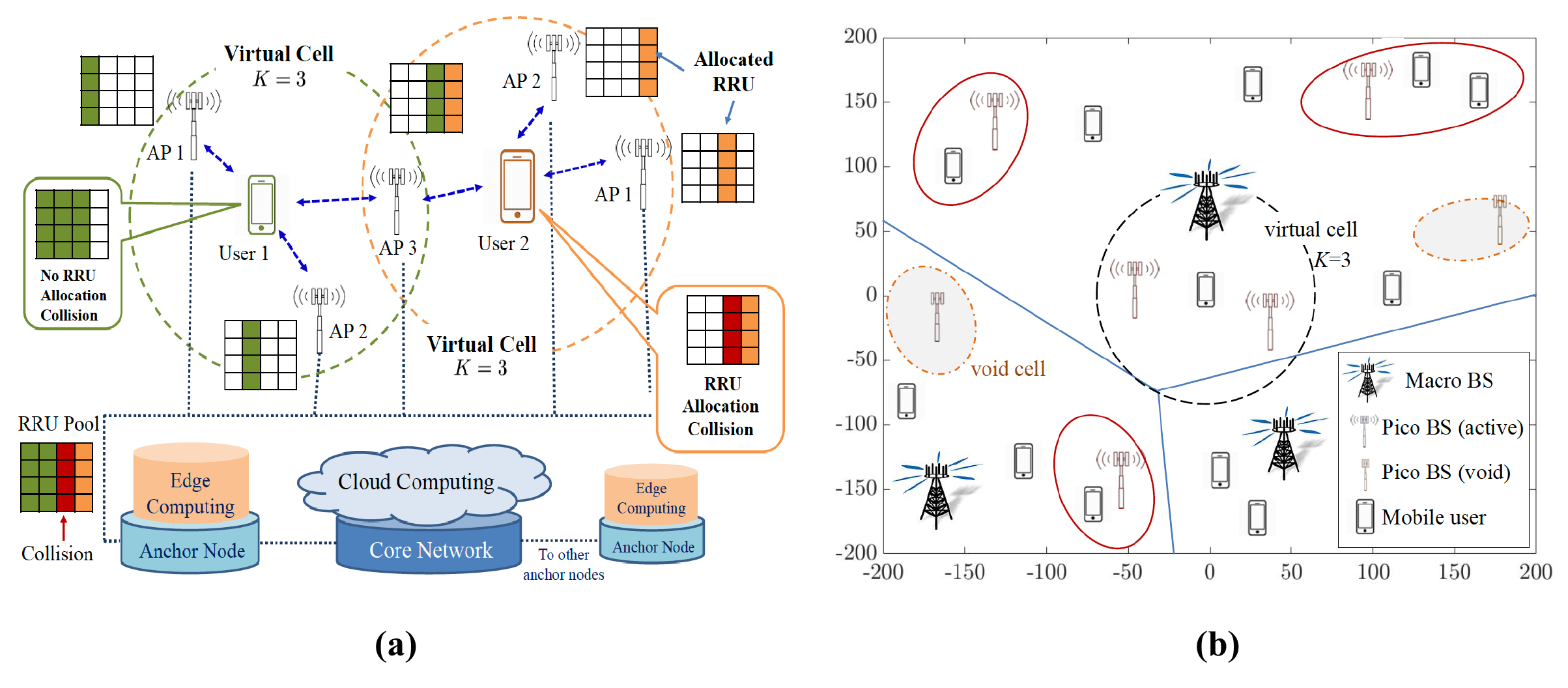}
	\caption{(a) Two URLLC users and their virtual cell with three APs: The two virtual cells share AP 3 and there is RRU (radio resource unit) allocation collision on User 2. (b) The void cell phenomenon after using the PMCA scheme. Note that all the three APs in a virtual cell connect to the same anchor node.}
	\label{Fig:SystemModel}
\end{figure*}

Although these aforementioned works and many others in the literature provide a good study on how to achieve URLLC and use it as a constraint to optimize the single-cell performance by using the closed-loop communication and retransmissions, they cannot reveal a good network-wide perspective  on how interferences from other cells and user/cell association schemes impact the URLLC performance of cellular systems. This paper aims to exploit the URLLC performances when open-loop communication is adopted in a large-scale cellular network. On account of open-loop communication, there is no feedback of channel state information in the network so that the advantages of multi-antenna transmission cannot be exploited in the network. Accordingly, all the analyses in this paper are conducted according to the assumption that all APs and users are equipped with a single antenna and they thus accomplish the following contributions: 
\begin{itemize}
	\item To enhance the communication reliability between users and APs, we propose the proactive multi-cell association (PMCA) scheme in which each user proactively associates with multiple nearby APs and adopts open-loop communication to send URLLC messages to the APs associated with it as well as receive URLLC messages from them. The PMCA scheme is completely different from any prior association schemes in the literature due to multi-cell association and transmission, which was also addressed in the recent standard of 3GPP for enhanced URLLC as an effective means to enhance communication reliability if compared with retransmission~\cite{3GPPR16}.
	\item The distribution of the number of the users associating with an AP in each tier is accurately derived, which is first found to the best of our knowledge. It importantly reveals that the void AP phenomenon exists in the PMCA scheme and needs to be considered in the analysis of ultra-reliable communications. 
	\item The uplink non-collision reliability of a user in the cell of an AP is found for the proactive open-loop communication and thereby we can characterize the uplink communication reliability of a virtual cell for the non-collaborative and collaborative AP cases in a low-complexity form. 
	\item From the analyses of the communication reliability in the uplink and downlink, the communication reliability is shown to be significantly influenced by the densities of the users and the APs and the number of the APs in a virtual cell. The PMCA scheme indeed improves the communication reliability of a user and make it achieve the target value of  $99.999\%$ when appropriately deploying APs for a given user density. 
	\item The uplink and downlink end-to-end delays between users and their anchor node are modeled and analyzed. We not only clarify the fundamental interplay among the delays, the number of the APs in a virtual cell, user and AP densities, but also show that achieving the target latency of one millisecond is certainly possible as long as the APs are deployed with a sufficient density for a given user density. 
\end{itemize}

\subsection{Paper Organization}\label{Subsec:Organization}
The rest of this paper is organized as follows. In Section \ref{Sec:SystemModel}, we first specify the system architecture of a HetNet for URLLC and then introduce the open-loop communication and propose the PMCA scheme. Section \ref{Sec:CommRel} models and analyzes the uplink and downlink communication reliabilities for the PMCA scheme and some numerical results are provided to validate the correctness and accuracy of the analytical results. In Section \ref{Sec:CommLatency}, the end-to-end latency problem for the open-loop communication and the PMCA scheme is investigated and some numerical results are also presented to evaluate the latency performance of the open-loop communication and the PMCA scheme. Finally, Section \ref{Sec:Conclusion} concludes our findings in this paper.

\section{System Model and Assumptions}\label{Sec:SystemModel}

\begin{table*}
	\caption {Notation of Main Variables, Symbols, and Functions}
	\centering\footnotesize
	\label{Tab:Notation}
	\begin{tabular}{ |c|c||c|c|}
		\hline
		Symbol & Meaning  & Symbol & Meaning \\ 
		\hline
		$\Phi_{m}$ & Set of tier-$m$ APs & $\gamma^{ul}_K$ & Uplink SIR of the $k$th AP in $\mathcal{V}_K$ \\
		$\lambda_m$ &  Density of $\Phi_m$ & $\eta^{ul}_K$ ($\eta^{dl}_K$) & Uplink (Downlink) communication reliability \\ 
		$P_m$ & Transmit power of the tier-$m$ APs & $\eta^{dl}_k$ &  Downlink reliability of  the $k$th AP in $\mathcal{V}_K$\\ $A_{m,i}$ & AP $i$ in the $m$th tier and its location & $\mathds{1}(\mathcal{E})$ & Indicator function of event $\mathcal{E}$ \\ $K$ &  Number of APs in a virtual cell & $\mathcal{L}_Z(s)$ & Laplace transform of random variable $Z>0$ \\  $\mathcal{V}_K$ &  Virtual cell with $K$ APs & $Q_k$ &  Transmit power of the $k$th AP in $\mathcal{V}_K$\\ $V_k$  & The $k$th AP in $\mathcal{V}_K$ & $\theta$ & SIR threshold for successful decoding \\ $\mathcal{U}$  & Set of users & $H_k$ &  Downlink channel gain from $V_k$ to typical user \\$\mu$  & Density of users & $h_k$ & Uplink channel gain from typical user to $V_k$ \\   $U_j$ & User $j$ and its location & $q_j$ & Transmit power of user $U_j$ \\ $S_K$ & The $K$th-truncated shot signal process & $q_k$ & Transmit power of a user in $\mathcal{V}_k$ for AP $V_k$ \\
		$\|X-Y\|$ &  Euclidean distance between nodes $X$ and $Y$ & $\vartheta_m$ & Probability of an AP in $\mathcal{V}_K$ from $\Phi_m$ \\
		$p_{m,0}$ & The void probability of Tier-$m$ APs  & $D^{ul}$ ($D^{dl}$) & Uplink (Downlink) communication delay  \\
		$\alpha>2$ & Pathloss exponent & $D^{ul}_{ba}$ ($D^{dl}_{ba}$) & Uplink (Downlink) backhaul delay\\
		$\rho^{ul}$ & Uplink non-collision reliability of an AP  & $D^{ul}_{tr}$ ($D^{dl}_{tr}$) & Uplink (Downlink) transmission delay \\
		$\rho^{ul}_K$ & Uplink non-collision reliability of a user & $\xi$ & Short block-length packet size\\
		$\delta$ & Probability of a user selecting an RRU in $\mathcal{V}_K$ & $\tau$ & Duration of transmission\\
		$\omega_m$ & Tier-$m$ association bias & $B$ & Bandwidth\\
		$O_{m,i}$ & Void indicator of AP $i$ in the $m$th tier & $\epsilon$ & Decoding error probability\\
		\hline
	\end{tabular}
\end{table*}

In this paper, we consider an interference-limited planar HetNet in which there are two tiers of APs and the APs in the same tier are of the same type and performance. In particular, the APs in the $m$th tier form an independent homogeneous Poisson point process (PPP) of density $\lambda_m$ and they can be expressed as set $\Phi_m$ given by
\begin{align}
\Phi_m\defn\{A_{m,i}\in\mathbb{R}^2: i\in\mathbb{N}\},\,\, m=\{1,2\},
\end{align}
where $A_{m,i}$ denotes AP $i$ in the $m$th tier and its location. Without loss of generality, we assume the first tier consists of the macrocell APs and the second tier consists of the small cell APs. A macrocell AP has a much larger transmit power than a small cell AP, whereas the density of the macro AP is much smaller than that of the small cell APs. To effectively achieve URLLC in the HetNet, an anchor node which governs a number of nearby APs according to the geographical deployment of the APs is co-located with the edge/fog computing facilities, and a cloud radio access architecture comprised of a core network and a cloud is also employed in the HetNet. Macrocell APs and anchor nodes are connected to the core network which helps send complex computing tasks to the cloud for further data processing and management. In addition, all (URLLC) users in the HetNet also form an independent homogeneous PPP 
of density $\mu$ and they are denoted by set $\mathcal{U}$ as
\begin{align}
\mathcal{U}\defn\{U_j\in\mathbb{R}^2: j\in\mathbb{N}\},
\end{align} 
where $U_j$ stands for user $j$ and its location. An illustration of the system model depicted here is shown in Fig. \ref{Fig:SystemModel} (a), and the main notations used in this paper are summarized in Table \ref{Tab:Notation}. The open-loop communication technique is adopted in the HetNet, i.e., no (immediate) feedback between a AP and a user. In the following, we elaborate on the main idea of how to employ open-loop communication to achieve URLLC in the HetNet.

\subsection{Open-loop Communication and Proactive Multi-cell Association}\label{SubSec:OpelLoopCommMultiCellAss}

As mentioned in Section \ref{Sec:Introduction}, closed-loop communication fundamentally incurs more latency than open-loop communication owing to feedback. This point manifests that open-loop communication turns out to be the best solution to reducing latency from the receiver perspective because feedback-related communication latency is completely avoided. However, the reliability performance of wireless communications could be seriously weakened due to no feedback transmission in that  it cannot be improved by using the hybrid automatic repeat request (ARQ), a combination of high-rate forward error-correcting coding and ARQ error control, which is commonly used in closed-loop communication. This phenomenon reveals that there seemingly exists a tradeoff between latency and reliability in wireless communications. However, this tradeoff can be absolutely alleviated or tackled by ultra-reliable open-loop communication since closed-loop feedback hardly further benefits the reliability of a wireless channel with extremely high reliability. 

To create an ultra-reliable open-loop communication context for the users in the HetNet, the users are suggested to \textit{proactively} associate with multiple APs at the same time so that their communication reliability can be improved by spatial channel diversity and even boosted whenever the associated multiple APs are able to do joint decoding. This proactive multi-cell association approach leads to the concept of the \textit{virtual cell }of users, that is, each user seems to form its own virtual cell that encloses all the APs associated with it \cite{SCHHHSMC18} and an illustration of the virtual cell is shown in Fig. \ref{Fig:SystemModel}(a). Note that all the APs in the same virtual cell are assumed to be connected to the same anchor node for the consideration of modeling simplicity and how user mobility impacts the URLLC performance of a virtual cell due to handover between macro APs is not considered in this paper. All the radio resources in a virtual cell can be scheduled and allocated by an anchor node utilizing the edge/fog computing technology. Thus, letting users form their virtual cell (i.e., associate with multiple APs) has an advantage in largely reducing control signaling for frequent handovers between small cell APs, which leads to no handover latency in a virtual cell. However, a user should not associate with too many APs at the same time because the signaling overhead due to multi-AP synchronization could deteriorate the latency performance of its virtual cell. In addition, the means of short packet transmission will be employed in a virtual cell to further shorten the communication latency.

In the following subsection, we will clarify the fundamental interplay among reliability, latency, and multi-cell association by formally proposing the PMCA scheme and analyzing its related statistical properties. For the sake of simplicity,  the following analyses are conducted by assuming all users and APs are equipped with a single antenna rather than multiple antennas.  The reasons are specified as follows. As previously mentioned, open-loop communication that is adopted in a virtual cell does not require immediate feedback of channel state information (CSI), and thereby it cannot fully enjoy the advantages of multi-antenna transmission, such as diversity and beamforming. Although space-time coding for multi-antenna transmission can help achieve diversity, it is too complicated to be practically fulfilled in the short packet transmission that is needed for ultra-low latency communication. As such, the single-antenna transmission performance is pretty similar to the multi-antenna transmission performance in a virtual cell. Moreover, we would like to analyze whether or not the proposed open-loop communication and PMCA scheme can accomplish the practical requirements of URLLC even in the worst-case scenario of single-antenna transmission.  

\subsection{The PMCA Scheme and Its Related Statistics}\label{SubSec:ProMultiCellAsso}
Assume that each user in the network associates with $K$ APs by the following PMCA scheme. Let $V_k$ be defined as\footnote{For the sake of modeling simplicity, we do not consider the shadowing effect in \eqref{Eqn:kthStrongestAP} since it does not affect the following analyses of the communication reliability in Section \ref{Sec:CommRel} according to the random conservation property found in \cite{CHLLCW16}. Further more, we assume $w_m$ and $\lambda_m$ for $m\in\{1,2\}$ are sufficiently large such that each user can detect almost all the APs in the network and associate with at least $K$ APs by using the PMCA scheme in~\eqref{Eqn:kthStrongestAP}.}
\begin{align} 
V_k\defn \begin{cases}
\arg\max_{m,i:A_{m,i}\in\Phi}\{\frac{w_m}{\|A_{m,i}\|^{\alpha}}\}, k=1\\ \arg\max_{m,i:A_{m,i}\in\{\hat{\Phi}_{k-1}\}}\{\frac{w_m}{\|A_{m,i}\|^{\alpha}}\}, k>1
\end{cases},\label{Eqn:kthStrongestAP}
\end{align}
where $k\in\mathbb{N}_+$, $\Phi\defn \bigcup_{m=1}^2\Phi_m$, $\hat{\Phi}_{k-1}=\Phi\setminus \bigcup_{j=1}^{k-1}V_j$, $\alpha>2$ is the path-loss exponent, positive constant $w_m$ is the tier-$m$ cell association bias, and $\|X-Y\|$ denotes the Euclidean distance between nodes $X$ and $Y$. For a typical user located at the origin, $V_k$ is thus the $k$th biased nearest AP of the typical user by averaging out the channel fading gain effect on the user side. More specifically, $V_k$ denotes the $k$th nearest AP of this typical user if $w_m=1$ for all $m\in\{1,2\}$, whereas $V_k$ becomes the $k$th strongest AP of the typical user if $w_m=P_m$ where $P_m$ is the transmit power of the tier-$m$ APs. The $K$ APs associated with the typical user can be expressed as a set given by
\begin{align}
\mathcal{V}_K\defn\bigcup_{k=1}^K V_k,
\end{align}
which is called the virtual cell of the typical user.  

\begin{figure*}[!t]
	\centering
	\setlength{\belowcaptionskip}{-2cm}
	\includegraphics[height=3.0in, width=1.05\textwidth]{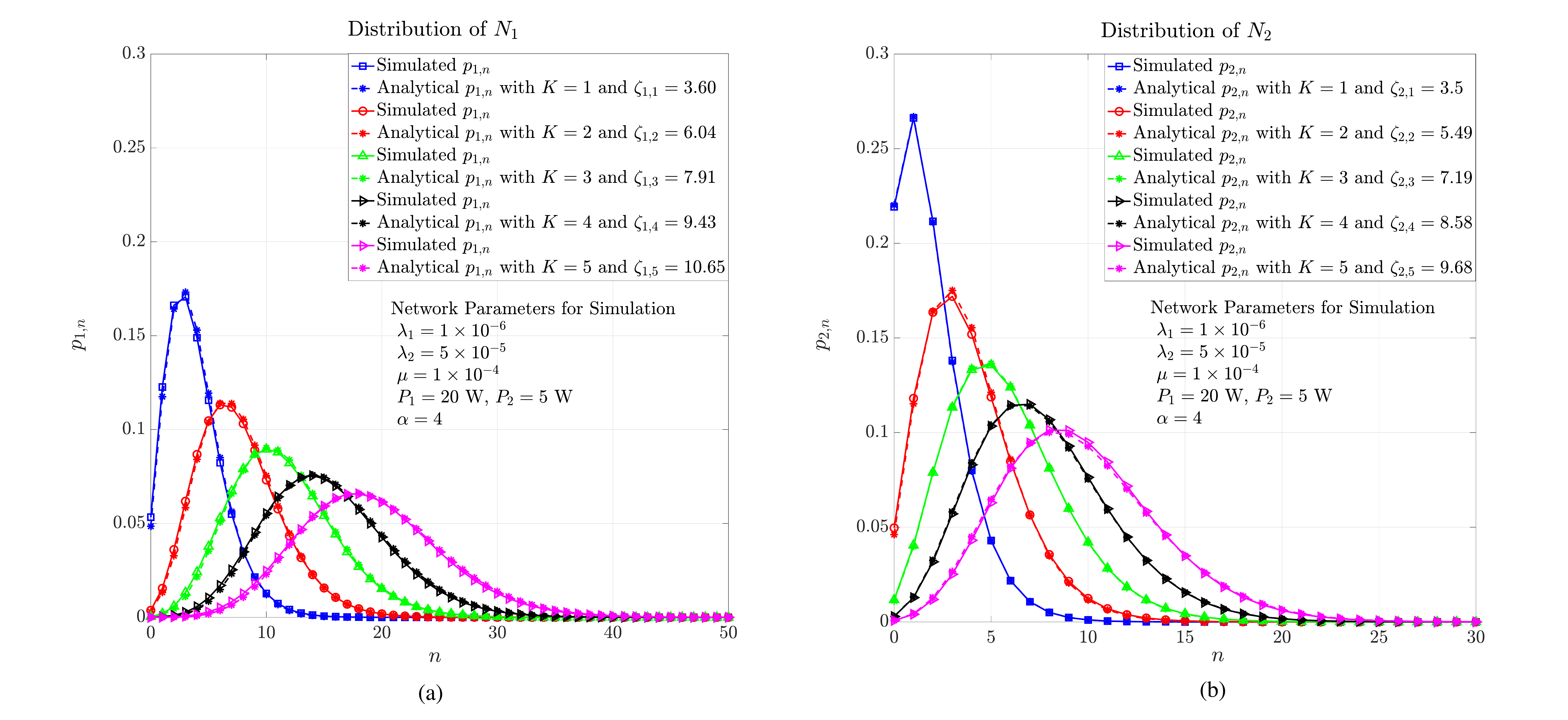}
	\caption{Simulation results of $p_{m,n}$: (a) $p_{1,n}$ for macro cell APs (b) $p_{2,n}$ for small cell APs.}
	\label{Fig:DisNumUsers}
\end{figure*}

According to \cite{CHLLCW1502}\cite{CHLKLF16}, the distribution of the number of the users associating with an AP is found for the single-cell association scheme. The method of deriving it cannot be directly applied to the case of the PMCA scheme because the cells of the APs are no longer disjoint in the multi-cell association case. Nonetheless, the idea behind the method is fairly helpful for us to derive the distribution of the number of the users within the cell of an AP for the PMCA scheme, as shown in the following lemma.
\begin{lemma}\label{Lem:PMFUserAssKAPs}
Suppose each user in the network adopts the PMCA scheme in \eqref{Eqn:kthStrongestAP} to associate with $K$ APs in the network. Let $N_m$ denote the number of the users associating with a tier-$m$ AP and its distribution (i.e., $p_{m,n}\defn\mathbb{P}[N_m=n]$) can be semi-analytically approximated as
\begin{align}\label{Eqn:PMFUserAssKAPs}
p_{m,n}\approx& \frac{\Gamma(n+\zeta_{m,K})}{n!\Gamma(\zeta_{m,K})} \left(\frac{K\mu}{\zeta_{m,K}\widetilde{\lambda}_m}\right)^n\nonumber\\
&\times\left(1+\frac{K\mu}{\zeta_{m,K}\widetilde{\lambda}_m}\right)^{-(n+\zeta_{m,K})},
\end{align}
where $\Gamma(x)\defn \int_{0}^{\infty} t^{x-1}e^{-t}\dif t$ for $x>0$ is the Gamma function, $\zeta_{m,K}>0$ is a positive constant that needs to be determined by the real numerical data of $p_{m,n}$, and $\widetilde{\lambda}_m\defn \sum_{i=1}^{2}w_i^{\frac{2}{\alpha}}\lambda_i/w^{\frac{2}{\alpha}}_m$.
\end{lemma}
\begin{IEEEproof}
See Appendix \ref{App:PMFUserAssKAPs}.
\end{IEEEproof}
To validate the correctness and accuracy of $p_{m,n}$ in \eqref{Eqn:PMFUserAssKAPs}, we adopt the network parameters for a two-tier HetNet shown in Fig. \ref{Fig:DisNumUsers} to numerically simulate $p_{m,n}$ for $K\in\{1,2,\ldots,5\}$. As can be seen in Fig. \ref{Fig:DisNumUsers}, the simulated results of $p_{m,n}$ accurately coincide with its corresponding analytical results of $p_{m,n}$ found in \eqref{Eqn:PMFUserAssKAPs}. Since the approximated result of $p_{m,n}$ in Lemma \ref{Lem:PMFUserAssKAPs} is numerically validated, there are some important implications that can be drawn. First, we can learn that the average number of the users associating with a tier-$m$ AP is $K\mu/\widetilde{\lambda}_m$ and this means the average cell size of a tier-$m$ AP is $K/\widetilde{\lambda}_m$ \cite{CHLKLF16,DSWKJM13}. In other words, the average cell size of an AP increases $K$ times as users associate with $K$ APs. Second, for $K=1$ users only associate with a single AP so that the cells of the APs do not overlap and the entire network area consists of weighted Voronoi-tessellated cells. For $K>1$, the cells of the APs may overlap in part and the cell sizes of the APs and the numbers of the users associating with the APs are no longer completely independent. Third, the probability that a tier-$m$ AP is not associated with any users, referred to as the tier-$m$ void probability, can be found as
\begin{align}\label{Eqn:Tier-mVoidProb}
p_{m,0}=\left(1+\frac{K\mu}{\zeta_{m,K}\widetilde{\lambda}_m}\right)^{-\zeta_{m,K}}.
\end{align}
For a dense cellular network with a moderate user density, this void probability may be so large that the void APs could be a considerable amount in the network. For example, we use the network parameters for simulation in Fig. \ref{Fig:DisNumUsers} to find the void probabilities $p_{1,0}=0.03$ and $p_{2,0}=0.2$ for $K=3$ and the void probability of small cell APs is actually not small at all (there are 20$\%$ of the small cell APs that are void.). Thus, such a void cell phenomenon for the PMCA scheme, as illustrated in Fig. \ref{Fig:SystemModel} (b), must be considered in the interference model \cite{CHLLCW1502,CHLLCW16} when the user density is not very large if compared with the density of the small cell APs. 

\subsection{The Truncated Shot Signal Process in a Virtual Cell}
\begin{figure}[!t]
	\centering
	\includegraphics[width=1.0\linewidth, height=2.5in]{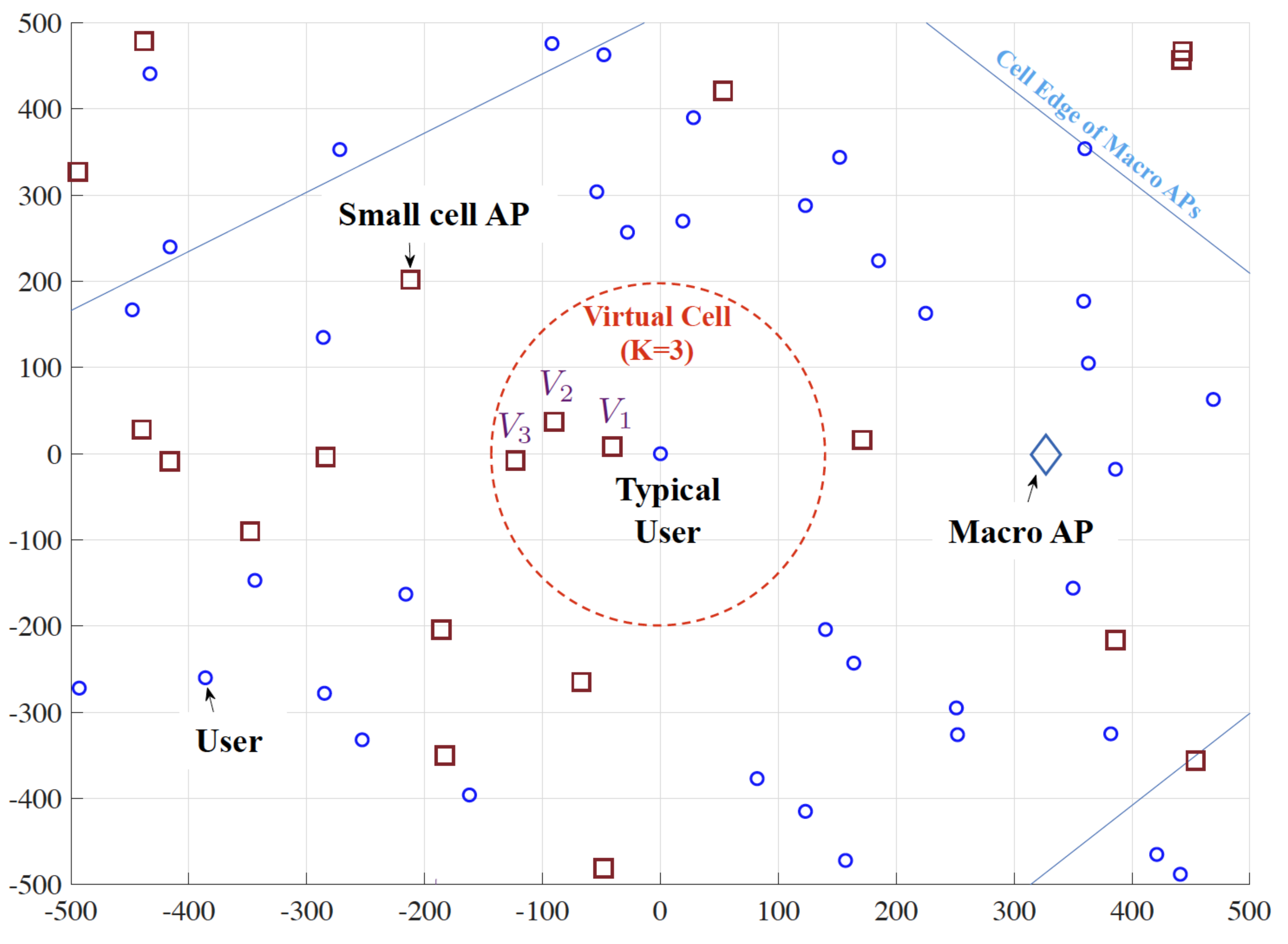}
	\caption{An illustrative example of the $K$th truncated shot signal process in a virtual cell for $K=3$. In the figure, the typical user form a virtual cell by associating with the first three nearest small cell APs using the PMCA scheme in~\eqref{Eqn:kthStrongestAP}, i.e., $V_1$, $V_2$, $V_3$. For this example, we have the 3rd truncated shot signal process $S_3=\sum_{k=1}^{3} H_k\|V_k\|^{-\alpha}$, which can be referred to as the sum of the (random) received powers from all $V_k$'s when all the small cell APs are transmitting with unit power.}
	\label{Fig:TrunShotSigProc}
\end{figure}

As the PMCA scheme and the virtual cell of a user introduced in Section \ref{SubSec:ProMultiCellAsso}, we define the $K$th-truncated shot signal process of the virtual cell of the typical user as follows\footnote{When $K$ goes to infinity, $S_{\infty}\defn \lim_{K\rightarrow\infty} S_K$ is traditionally referred to as (complete) Poisson shot noise process \cite{SBLMCT90,CHLJGA12} since it contains weighted signal powers in a Poisson field of transmitters. Since $S_K$ only contains the signals emitted from the first $K$ weighted nearest transmitters in the network, it is called the $K$th-truncated shot signal process.}:
\begin{align}\label{Eqn:TrunKthShotNoiseProc}
S_K\defn  \sum_{k=1}^{K} H_kW_k\|V_k\|^{-\alpha},
\end{align}
where $V_k\in\mathcal{V}_K$ is already defined in~\eqref{Eqn:kthStrongestAP}, $H_k$ denotes the fading channel gain from $V_k$ to the typical user, $W_k\in\{w_1,w_2\}$ is the cell association bias of AP $V_k$, and it is a non-negative random variable (RV) associated  with $V_k$. We call $S_K$ the $K$th-truncated shot signal process because it only captures the cumulative effect at the typical user of the $K$ random shocks from the $K$ different random locations (i.e., $V_1,\ldots, V_K$), and $H_kW_k\|V_k\|^{-\alpha}$ can be viewed as the impulse function of AP $V_k$ that gives the $H_kW_k$-weighted attenuation of the transmit power of $V_k$ in space. An illustrative example of the $K$th truncated shot signal process in a virtual for $K=3$ is visually demonstrated in Fig.~\ref{Fig:TrunShotSigProc} where the densities of the users, macro APs, and small cell APs are $50$ users/km$^2$, $1.0$ APs/km$^2$, and $10$ APs/km$^2$, respectively. In the figure, the typical user adopts the PMCA scheme in~\eqref{Eqn:kthStrongestAP} with $w_1=w_2=1$ to form its virtual cell by associating with the first 3 nearest (small cell) APs in the HetNet. Consequently, there is the $3$rd-truncated shot signal process in the virtual, i.e., $S_3=\sum_{k=1}^{3} H_k \|V_k\|^{-\alpha}$, which can be interpreted as the sum of the (desired) received signal powers from all $V_k$'s in the virtual cell while all the small cell APs are transmitting with unit power.

Let $\mathcal{L}_Z(s) \defn \mathbb{E}[\exp(-sZ)]$ denote the Laplace transform of a non-negative RV $Z$ for $s>0$ and some statistical results regarding $S_K$ are presented in the following theorem.
\begin{theorem}\label{Thm:LapTramsSK}
Assume all the $H_k$'s of the $K$th-truncated shot signal process in \eqref{Eqn:TrunKthShotNoiseProc} are i.i.d. exponential RVs with unit mean\footnote{In this paper, we consider the HetNet is in a rich-scattering environment so that all the fading channel gains of the wireless links in the HetNet can be assumed to be independent due to very weak spatial correlation between them.}, i.e., $H_k\sim\exp(1)$. If we define $S_{-K}\defn S_{\infty}-S_K$ and $S_{\infty}\defn \lim_{K\rightarrow\infty}S_K$, then the Laplace transform of $S_{-K}$ can be explicitly found as
\begin{align}\label{Eqn:LapTransS-K}
\mathcal{L}_{S_{-K}}(s) =& \frac{(\pi\widetilde{\lambda})^K}{(K-1)!} \int_{0}^{\infty} y^{K-1}\nonumber\\
&\times \exp\left\{-\pi\widetilde{\lambda} y\left[1+\ell\left(sy^{-\frac{\alpha}{2}} ,\frac{2}{\alpha}\right)\right]\right\}\dif y,
\end{align}
where $\widetilde{\lambda}\defn \sum_{m=1}^{2}w^{\frac{2}{\alpha}}_m\lambda_m$, $\vartheta_m\defn \mathbb{P}[W_k=w_m]= w^{\frac{2}{\alpha}}_m\lambda_m/\widetilde{\lambda}$ is the probability that a user associates with a tier-$m$ AP, $\ell(y,z)$ for $y,z\in\mathbb{R}_+$ is defined as
\begin{align}\label{Eqn:EllFun}
\ell(y,z)\defn\frac{y^{z}}{\mathrm{sinc}(z)}-\int_{0}^{1} \frac{y}{y+t^{\frac{1}{z}}}\dif t,
\end{align}
and $\mathrm{sinc}(x)\defn \frac{\sin(\pi x)}{\pi x}$. For the Laplace transform of $S_K$, it can be explicitly found as
\begin{align}\label{Eqn:LapTransSK}
\mathcal{L}_{S_K}(s) =& \exp\left[-\frac{\pi\widetilde{\lambda}s^{\frac{2}{\alpha}}}{\mathrm{sinc}(2/\alpha)}\right]\frac{(\pi\widetilde{\lambda})^K}{(K-1)!}\int_{0}^{\infty} y^{K-1}\nonumber\\
&\times  \exp\left\{\pi\widetilde{\lambda}y\left[1+\ell\left(sy^{-\frac{\alpha}{2}},\frac{2}{\alpha}\right)\right]\right\}\dif y.
\end{align}
In addition, the upper bound on $\mathbb{P}[S_K\geq y]$ can be found as
\begin{align}\label{Eqn:LowBoundSK}
\mathbb{P}[S_K\geq y] \leq  1-\prod_{k=1}^{K}\left[1-\mathcal{L}_{Y^{\frac{\alpha}{2}}_k}\left(\frac{2ky}{K(K+1)}\right)\right],
\end{align}
where $Y_k\sim\text{Gamma}(k,\pi\widetilde{\lambda})$ is a Gamma RV with shape parameter $k$ and rate parameter $\pi\widetilde{\lambda}$.
\end{theorem}
\begin{IEEEproof}
See Appendix \ref{App:ProofLapTransSK}.
\end{IEEEproof}
The above results of Laplace transform in Theorem \ref{Thm:LapTramsSK} indicate that in general the closed-form results of $\mathcal{L}_{S_{-K}}$ and $\mathcal{L}_{S_K}$ are unable to be obtained except in some special cases. For instance, letting $s=\varphi y^{\frac{\alpha}{2}}$ for $\varphi>0$ and $\mathcal{L}_{y^{\frac{\alpha}{2}}S_{-K}}(\varphi)$ can be shown as
\begin{align}\label{Eqn:WeighedSK}
\mathcal{L}_{y^{\frac{\alpha}{2}}S_{-K}}(\varphi) &=\frac{(\pi\widetilde{\lambda})^K}{(K-1)!} \int_{0}^{\infty} y^{K-1}e^{-\pi\widetilde{\lambda} y\left[1+\ell\left(\varphi ,\frac{2}{\alpha}\right)\right]}\dif y\nonumber\\
&=\left[1+\ell\left(\varphi,\frac{2}{\alpha}\right)\right]^{-K}.
\end{align}
Nonetheless, we can still resort to some numerical techniques to evaluate the Laplace transforms of $S_{-K}$ and $S_K$ and the distributions of $S_{-K}$ and $S_K$ by numerically evaluating the inverse Laplace transform of $S_{-K}$ and $S_K$. In addition, we are still able to understand the distribution behaviors of $S_K$ from the closed-form upper bound on $\mathbb{P}[S_K\geq y]$. Theorem \ref{Thm:LapTramsSK} plays an important role in the following analysis of the communication reliability that will be defined in the following section.

\section{Communication Reliability Analysis for Proactive Multi-cell Association}\label{Sec:CommRel}
In this section, we would like to exploit the fundamental performances and limits of the communication reliability of users in the uplink and the downlink when the PMCA scheme is employed in the HetNet. We assume that the orthogonal frequency division multiple access is adopted in the HetNet and the communication reliability analyses are proceeded in accordance with how the radio resource blocks (RB) in the cell of each AP are requested by a user in the uplink and allocated by an AP in the downlink. We will first specify how users access the RBs of an AP and then propose and analyze the uplink communication reliability. Afterwards, we will continue to study the communication reliability in the downlink case.  

\subsection{Analysis of Uplink Communication Reliability}\label{Subsec:AnaUplinkCommRel}
According to the PMCA scheme and the virtual cell of a user specified in Section \ref{SubSec:ProMultiCellAsso}, our interest here is to study how likely a user is able to successfully access  available RBs of an AP and then send its message to the $K$ APs in its virtual cell through the open-loop communication. To establish the uplink access from a user to the $K$ APs, we propose the following PMCA-based radio resource allocation scheme for  uplink open-loop communication: 
\begin{itemize}
	\item To make a user have good uplink connections, the user forms its virtual cell by associating with its first $K$ nearest APs. Thus, all the cell association biases in \eqref{Eqn:kthStrongestAP} are unity, i.e., $w_m=1$ for all $m\in\{1,2\}$. 
	\item Each radio RB serves as the basic unit while scheduling radio resources. Multiple radio RBs in a single time slot are mapped to a single (virtual) radio resource unit (RRU) for transmitting a message. Users are allowed to transmit one message in each time slot. 
	\item Due to lack of CSI of each AP in the virtual cell\footnote{ Note that the open-loop communication can be applied to frequency-division duplex (FDD) systems and  time-division duplex (TDD) systems. As such, it is not necessary to designate the HetNet in the paper to adopt either FDD or TDD. Our goal in this paper is to delve how to achieve URLLC in the HetNet under the CSI-free scenario.}, a user proactively allocates the radio resource in a distributed manner, that is, it has to randomly select RRUs for the $K$ APs in its virtual cell\footnote{In a virtual cell, a user is not suggested to adopt carrier-sense multiple access (CSMA) protocols to gain the radio resource since the channel access latency induced by CSMA is too large to be satisfied by the ultra-low latency requirement of URLLC.}.
\end{itemize}

Since each user has to randomly select the uplink RRUs in its virtual cell without considering how other users select their uplink RRUs, multiple users in the cell of an AP could select the same RRUs, which leads to transmission collisions as indicated in Fig. 1(a). If users associated with the same AP select the same RRU, then they fail to upload their message to their APs owing to transmission collisions between them. The probability that there is no uplink collision in the virtual cell, referred to as the uplink non-collision reliability, is found in the following lemma.
\begin{lemma}\label{Lem:NoUplinkColProb}
Suppose a user adopts the PMCA scheme in  \eqref{Eqn:kthStrongestAP} to form its virtual cell with $K$ APs. If the probability that the user selects any one of the RRUs for each AP in its virtual cell is $\delta\in(0,1)$, then the uplink non-collision reliability of each AP in its virtual cell is found as
\begin{align}\label{Eqn:NonCollProbEachAP}
\rho^{ul} = \sum_{m=1}^{2}\vartheta_m\sum_{n=1}^{\infty}p_{m,n}(1-\delta)^{n-1}.
\end{align}
Hence, the uplink non-collision reliability of the user in its virtual cell is
\begin{align}\label{Eqn:NonColProb}
\rho^{ul}_K= 1-\left[1-\rho^{ul}\right]^K.
\end{align}

\end{lemma}
\begin{IEEEproof}
See Appendix \ref{App:ProofNoUplinkColProb}.
\end{IEEEproof}

\noindent Lemma \ref{Lem:NoUplinkColProb} reveals that the uplink non-collision reliability of each AP is mainly influenced by $K$ and $\delta$, e.g., it decreases whenever $p_{m,n}$ decreases by increasing $K$ and/or $\delta$ decreases\footnote{In general, $\vartheta_m$ does not have a significant impact on $\rho^{ul}_K$ in that usually $P_2\gg P_1$ as well as $\lambda_1\ll \lambda_2$ and these two condition leads to $\vartheta_1\ll\vartheta_2$ in most of practical cases}; thereby, fewer APs in the virtual cell and/or more radio resources may significantly improve the uplink non-collision reliability. Note that the uplink non-collision reliability of a user may not always increase as $K$ increases since $\rho^{ul}_K\approx K\rho^{ul}$ for $\rho^{ul}\ll 1$ and increasing $K$ in this situation may not increase $\rho^{ul}_K$ because $\rho^{ul}$ decreases in this case. In addition to the uplink collision problem happening to APs, whether a user is able to successfully send its messages to at least one AP in its virtual cell also depends upon all the communication link statuses in the virtual cell. Let $\gamma^{ul}_k$ denote the uplink signal-to-interference ratio (SIR) from a typical user located at the origin to the $k$th AP in the virtual cell, and it can be expressed as
\begin{align}\label{Eqn:UplinkSIR}
\gamma^{ul}_k \defn \frac{h_kq_k\|V_k\|^{-\alpha}}{\sum_{j:U_j\in\mathcal{U}_{a}}h_jq_j\|V_k-U_j\|^{-\alpha}},
\end{align}
where $h_k$ denotes the uplink fading channel gain from the typical user to AP $V_k$,   $q_k$ is the transmit power used by the typical user for AP $V_k$, $q_j$ is the transmit power of user $U_j$, $h_{jk}$ is the uplink fading channel gain from $U_j$ to $V_k$, and $\mathcal{U}_a\subseteq \mathcal{U}$ represents the set of the actively transmitting users using the same RRU as the typical user. All uplink fading channel gains are assumed to be i.i.d. exponential RVs with unit mean, i.e., $h_k,h_{jk}\sim\exp(1)$ for all $k,j\in\{1,\ldots,K\}$.  

According to \eqref{Eqn:UplinkSIR}, we can consider two cases of non-collaborative and collaborative APs to define the uplink communication reliability in a virtual cell. The case of non-collaborative APs corresponds to the situation in which all APs in the virtual cell are not perfectly coordinated so that they cannot do joint transmission and reception, whereas when all APs in the virtual cell are perfectly coordinated so that they are able to collaboratively do joint transmission and reception corresponds to the case of collaborative APs. For the case of non-collaborative APs, the uplink communication reliability is defined as the probability that a message sent by a user in a virtual cell is successfully received by at least one non-collision AP in the virtual cell, and it can be expressed as 
\begin{align}\label{Eqn:UplinkCommRel}
\eta^{ul}_K \defn \mathbb{P}\left[\max_{k\in\{1,\ldots,K\}}\{\gamma^{ul}_k\mathds{1}(V_k\in \mathcal{V}^{nc}_K)\}\geq \theta\right], 
\end{align}
where $\mathds{1}(\mathcal{A})$ is the indicator function that is unity if event $\mathcal{A}$ is true and zero otherwise, $\theta>0$ is the SIR threshold for successful decoding, and $\mathcal{V}^{nc}_K\subseteq\mathcal{V}_K$ is the subset of the APs without collision in 
set $\mathcal{V}_K$.  For the case of collaborative APs, the uplink communication reliability is defined as
\begin{align}\label{Eqn:UplinkCommRelCoop}
\eta^{ul}_K \defn \mathbb{P}\left[\frac{S^{ul}_K}{\sum_{j:U_j\in\mathcal{U}_{a}\setminus\mathcal{V}_K}h_jq_j\|V_k-U_j\|^{-\alpha}}\geq\theta \right],
\end{align}
where $S^{ul}_K \defn \sum_{k:V_k\in\mathcal{V}^{nc}_K} h_kq_k\|V_k\|^{-\alpha}$. Namely, $\eta^{ul}_K$ in \eqref{Eqn:UplinkCommRelCoop} is the probability that a uplink message is successfully received by at least one non-collision AP in the virtual cell: If there are at least two non-collision APs in the virtual cell, they can jointly decode the message. Otherwise, only one non-collision AP can decode it.\footnote{For the sake of analytical tractability, we consider that non-coherent signal combing happens among all the non-collision APs in the virtual cell even though such a combining leads to a suboptimal SIR performance.}

The analytical results of \eqref{Eqn:UplinkCommRel} and \eqref{Eqn:UplinkCommRelCoop} are summarized in the following theorem.
\begin{theorem}\label{Thm:UplinkCommRel}
Suppose each user employs the PMCA scheme in \eqref{Eqn:kthStrongestAP} to form its virtual cell with $K$ APs. If all the $K$ APs in the virtual are unable to collaborate, the uplink communication reliability defined in \eqref{Eqn:UplinkCommRel} is approximated by
\begin{align}\label{Eqn:UplinkCommRelRel}
\eta^{ul}_K \approx 1-\prod_{k=1}^K \left\{1-\rho^{ul}\left(1+\frac{\delta \theta^{\frac{2}{\alpha}}(1-p_0)}{\mathrm{sinc}(2/\alpha)}\right)^{-k}\right\},
\end{align}
where $p_0\defn p_{m,0}$ that is given in \eqref{Eqn:Tier-mVoidProb} with $w_m=1$ for $m\in\{1,2\}$ and $\rho^{ul}$ is given in \eqref{Eqn:NonCollProbEachAP}. For the case of $K\rightarrow\infty$, $\eta^{ul}_{\infty}\defn\lim_{K\rightarrow\infty}\rho^{ul}_K$ can be approximately found as
\begin{align}\label{Eqn:UplinkCommRelManyAPs}
\eta^{ul}_{\infty}\approx 1-\exp\left[- \frac{\rho^{ul}}{\delta\theta^{\frac{2}{\alpha}}}\mathrm{sinc}\left(\frac{2}{\alpha}\right) \right].
\end{align}
When all the $K$ APs in the virtual cell are able to collaborate to jointly decode the uplink message, $\eta^{ul}_K$ in \eqref{Eqn:UplinkCommRelCoop} can be upper bounded by
\begin{align}\label{Eqn:UplinkRelCop}
\eta^{ul}_K \leq & \rho^{ul}\bigg\{1-\prod_{k=1}^{K}\bigg[1-\bigg(1+\frac{\delta(1-p_0)}{ \mathrm{sinc}(2/\alpha)}\nonumber\\
&\times\bigg(\frac{2k\theta}{K(K+1)}\bigg)^{\frac{2}{\alpha}}\bigg)^{-k}\bigg]\bigg\}.
\end{align}
\end{theorem}
\begin{IEEEproof}
See Appendix \ref{App:ProofUplinkCommRel}.
\end{IEEEproof}
\noindent From the results in Theorem \ref{Thm:UplinkCommRel}, the uplink reliability $\eta^{ul}_K$ in \eqref{Eqn:UplinkCommRelRel} monotonically increases as $K$ increases even though increasing $K$ makes $p_0$ reduce and it thereupon reduces the number of the void cells and induces more interference. However, $\eta^{ul}_K$ suffers from the diminishing returns problem as $K$ increases so that associating with too many APs may not be an efficient means to significantly improve $\eta_K$ for a user. In particular, \eqref{Eqn:UplinkCommRelRel} can be used to obtain the following result
\begin{align}
\frac{1-\eta^{ul}_K}{1-\eta^{ul}_{K-1}} = 1-\rho^{ul} \left(1+\frac{\delta\theta^{\frac{2}{\alpha}}}{\mathrm{sinc}(2/\alpha)}\right)^{-K},
\end{align}
which indicates $(1-\eta^{ul}_K)/(1-\eta^{ul}_{K-1})\approx 1$ as $K\gg 1$ and we thus know $\eta^{ul}_K/\eta_{K-1}\approx 1$ for large $K$, i.e., the diminishing returns problem occurs. According to \eqref{Eqn:UplinkCommRelRel}-\eqref{Eqn:UplinkRelCop}, another two efficient approaches to boosting $\eta^{ul}_K$ are reducing the probability of scheduling each RRU in each cell and densely deploying APs in the HetNet, that is, we need small $\delta$ in that small $\delta$ suppresses the magnitude of the interference. For instance, if $\theta=6$ dB, $\alpha=4$, then $\eta^{ul}_{\infty}\approx 57.9\%$ for $\delta=0.5, \rho^{ul}=0.9$ and $\eta^{ul}_{\infty}\approx 98.96\%$ for $\delta=0.1, \rho^{ul}=0.95$. Note that $\eta^{dl}_{\infty}$ in \eqref{Eqn:UplinkCommRelManyAPs} characterizes the fundamental limit of the uplink communication reliability if all APs cannot collaborate in the uplink and it can be used to evaluate whether the PMCA and resource allocation schemes can achieve some target value of $\eta^{ul}_K$. If $\theta=6$ dB and $\alpha=4$, for example, we require $\delta\leq 5\%$ in order to achieve $\eta^{ul}_{\infty}\geq 99.999\%$. In other words, the target reliability $99.999\%$ is not able to be achieved by the PMCA scheme if $\delta>5\%$. Note that $\eta^{ul}_K$ in \eqref{Eqn:UplinkCommRelCoop} is certainly larger than that in \eqref{Eqn:UplinkCommRel} and the upper bound on $\eta^{ul}_K$ in \eqref{Eqn:UplinkRelCop} may be greater than the result in \eqref{Eqn:UplinkCommRelManyAPs}. In addition, we would like to point out that a virtual cell with $K$ APs can support uplink short packet transmission when $\eta^{ul}_K$ is higher than the target reliability because short packet transmission suffers from the problem of degraded transmission reliability and efficiency and such a problem is significantly mitigated by a high value of $\eta^{ul}_K$. These aforementioned observations will be numerically validated in Section \ref{SubSec:NumResCommRel}.

\begin{table*}[!t]
	\centering\small
	\caption{Network Parameters for Simulation~\cite{CSCSCYTQYL19,WYGDTKYP14}}\label{Tab:SimPara}
	\begin{tabular}{|c|c|c|}
		\hline Parameter $\setminus$ AP Type (Tier $m$)& Macrocell AP (1) & Small cell AP (2)\\ 
		\hline Transmit Power $P_m$ (W) & 20 & 5\\
		\hline User Density $\mu$ (users/km$^2$) &\multicolumn{2}{c|}{$50$} \\ 
		\hline AP Density $\lambda_m$ (APs/km$^2$) & $1.0$ & $[0.2\mu, \mu] $  \\ 
		\hline RRU Selection Probability $\delta$ & \multicolumn{2}{c|}{0.05} \\
		\hline Path-loss Exponent $\alpha$ & \multicolumn{2}{c|}{4} \\  
		\hline Tier-$m$ Association Bias $w_m$ (Uplink, Downlink) & \multicolumn{2}{c|}{$(1,P_m)$}  \\ 
		\hline Packet Size $\xi$ (bytes) & \multicolumn{2}{c|}{8, 32, and 64}  \\ 
		\hline Duration of Transmission $\tau$ (ms) & \multicolumn{2}{c|}{0.05}  \\ 
		\hline Bandwidth $B$ &\multicolumn{2}{c|}{20 MHz} \\ 
		\hline Decoding Error Probability $\epsilon$ &\multicolumn{2}{c|}{$2\times10^{-8}$} \\ 
		\hline SIR Threshold $\theta$ &\multicolumn{2}{c|}{$\exp\left[\frac{\xi\ln 2}{\tau B} + \frac{Q^{-1}_G(\epsilon)}{\sqrt{\tau B}} \right]-1$} \\ 
		\hline
	\end{tabular} 
\end{table*}

\subsection{Analysis of Downlink Communication Reliability}\label{SubSec:AnaDLCommRel}
In this subsection, we would like to study the downlink communication reliability of users in their virtual cells. Note that the uplink transmission and the downlink transmission in a virtual cell are independent and there is thus no order between them. To establish the benchmark performance, we assume that the frequency reuse factor in this cellular network is unity (i.e., all APs share the entire available frequency band) so that we can evaluate the downlink communication reliability in the worst-case scenario of interference. We also assume that each of the downlink RRUs of an AP is uniquely allocated to a user associating with the AP and users adopt the PMCA scheme to associate with the first $K$ strongest APs (i.e., $w_m=P_m$ in \eqref{Eqn:kthStrongestAP}). In the virtual cell of the typical user, the SIR of the link from the $k$th strongest AP to the typical user is defined as
\begin{align}\label{Eqn:Downlink-kthSIR}
\gamma^{dl}_k \defn \frac{H_kQ_k\|V_k\|^{-\alpha}}{\sum_{i:V_i\in\mathcal{V}_K\setminus V_k}H_iQ_i\|V_i\|^{-\alpha}+I^{dl}_K},
\end{align}
where $H_k$ denotes the downlink fading channel gain from $V_k$ to the typical user, $Q_k\in\{P_1,P_2\}$ is the transmit power of $V_k$, $H_{m,i}$ is also the downlink fading channel gain from $A_{m,i}$ to the typical user, $I^{dl}_K\defn \sum_{m,i:A_{m,i}\in\Phi\setminus\mathcal{V}_K}O_{m,i}H_{m,i}P_m\|A_{m,i}\|^{-\alpha}$, and $O_{m,i}\in\{0,1\}$ is a Bernoulli RV that is unity if $A_{m,i}$ is not void and zero otherwise. All $H_i$'s and $H_{m,i}$'s are assumed to be i.i.d. exponential RVs with unit mean. Note that $\mathbb{P}[O_{m,i}=1]=1-p_{m,0}$ and it can be found by using \eqref{Eqn:Tier-mVoidProb}. The term $I^{dl}_K$  in \eqref{Eqn:Downlink-kthSIR} is the interference from all non-void APs that are not in the virtual cell, whereas the term $\sum_{i:V_i\in\mathcal{V}_K\setminus V_k}H_iQ_i\|V_i\|^{-\alpha}$ in \eqref{Eqn:Downlink-kthSIR} is the intra-virtual-cell interference from other $K-1$ APs in the virtual cell if the $K-1$ APs in the virtual cell are not coordinated to avoid using the RRU used by the $k$th AP. This represents the worst case of the downlink SIR of the $k$th AP in the virtual cell. In this case, the downlink communication reliability of a virtual cell with $K$ APs is defined as\footnote{Due to open-loop communication, each AP does not have channel state information and thereby only the multi-AP (multi-channel) diversity can be exploited while the APs in a virtual cell are performing downlink CoMP. In the case of non-collaborative APs, the downlink transmission in a virtual cell succeeds as long as at least one AP can successfully transmit to the user in the virtual cell. The downlink communication reliability is thus defined as shown in \eqref{Eqn:DownConnReli}.}
\begin{align}\label{Eqn:DownConnReli}
\eta^{dl}_K\defn \mathbb{P}\left[\max_{k\in\{1,\ldots,K\}}\{\gamma^{dl}_k\}\geq \theta\right],
\end{align}
which is the probability that there is at least one AP in the virtual cell that can successfully transmit to the user in the virtual cell. 

When all the $K$ APs in the virtual cell can collaborate to eliminate the intra-virtual-cell interference, the downlink communication reliability can be simply written as
\begin{align}\label{Eqn:RedDownlinkSIRCoop}
\eta^{dl}_K = \mathbb{P}\left[\frac{\sum_{k=1}^{K}H_kQ_k\|V_k\|^{-\alpha}}{\sum_{m,i:A_{m,i}\in\Phi\setminus\mathcal{V}_K}O_{m,i}H_{m,i}P_m\|A_{m,i}\|^{-\alpha}}\geq\theta\right].
\end{align}
The explicit results of $\eta^{dl}_K$ defined in \eqref{Eqn:DownConnReli} and \eqref{Eqn:RedDownlinkSIRCoop} are found in the following theorem.
\begin{theorem}\label{Thm:DownlinkReliability}
Suppose all APs in a virtual cell are not coordinated so that there exists the intra-virtual-cell interference in the virtual cell. The downlink communication reliability in the case of non-collaborative APs defined in \eqref{Eqn:DownConnReli} is explicitly upper bounded by
\begin{align}\label{Eqn:UppBoundDLRel}
\eta^{dl}_K \leq 1-\prod_{k=1}^{K} \bigg\{ & 1-\left[1+\delta\ell\left(\theta,\frac{2}{\alpha}\right)\sum_{m=1}^{2}\vartheta_m(1-p_{0,m})\right]^{-k}\nonumber\\
&\bigg\},
\end{align}
where $\vartheta_m\defn P^{2/\alpha}_m\lambda_m/\widetilde{\lambda}$ and $\widetilde{\lambda}=\sum_{m=1}^{2}P^{\frac{2}{\alpha}}_m\lambda_m$. When $K$ goes to infinity, $\eta^{dl}_{\infty}\defn\lim_{K\rightarrow\infty} \eta^{dl}_K$ can be approximately found in closed form given by
\begin{align}\label{Eqn:BoundDownConnRel}
\eta^{dl}_{\infty} \approx  1-\exp\left[-\frac{\mathrm{sinc}\left(\frac{2}{\alpha}\right)}{\delta\theta^{\frac{2}{\alpha}}}\right].
\end{align}
For the case of collaborative APs, the upper bound on $\eta^{dl}_K$ in \eqref{Eqn:RedDownlinkSIRCoop} can be found as
\begin{align}\label{Eqn:RelDownLinkCoop}
\eta^{dl}_K \leq  1-\bigg\{ &1-\left[1+\delta\ell\left(\frac{\theta}{K^{\frac{\alpha}{2}+1}},\frac{2}{\alpha}\right)\sum_{m=1}^{2}\vartheta_m(1-p_{m,0})\right]^K\nonumber\\
&\bigg\}^{-K}.
\end{align}
\end{theorem}
\begin{IEEEproof}
See Appendix \ref{App:ProofDownlinkReliability}.
\end{IEEEproof}
\noindent From the results in Theorem \ref{Thm:DownlinkReliability}, we realize that increasing $K$ indeed improves $\eta^{dl}_K$ even though it reduces the tier-$m$ void probability $p_{m,0}$, yet it also suffers from the diminishing returns problem, like the uplink communication reliability. The tier-$m$ void probability $p_{m,0}$ also significantly impacts $\eta^{ul}_K$ when the number of the APs in a virtual cell is not large so that increasing the AP density improves $\eta^{dl}_K$ since it helps increase $p_{m,0}$. Moreover, $\delta$ can be interpreted as the probability that all APs statically allocate their RRUs with equal probability and it has to be small in order to achieve ultra-reliable communications. The result in \eqref{Eqn:BoundDownConnRel} that does not depend on the densities of the APs and users is the fundamental limit of the downlink communication reliability when all non-collaborative APs use different RRUs to transmit a message to the same user. It reveals whether ultra-reliable communications can be attained by the PMCA and resource allocation schemes. For example, if $\theta=6$ dB and $\alpha=4$, we need $\delta<0.05$ to achieve $\eta^{dl}_K\geq 99.999\%$, i.e., each RRU cannot be scheduled with a probability more than $5\%$ in this case. Otherwise, the PMCA scheme cannot successfully achieve the downlink communication reliability of $99.999\%$  no matter how many APs are in a virtual cell. Furthermore, $\eta^{dl}_K$ in  \eqref{Eqn:RelDownLinkCoop} for the case of collaborative APs is certainly higher than that in \eqref{Eqn:UppBoundDLRel} and it can also provide some insights into how to schedule resources and deploy APs in the HetNet so as to achieve the predesignated target value of $\eta^{dl}_K$. Likewise, a virtual cell with $K$ AP is able to support downlink short packet transmission when $\eta^{dl}_K$ is extremely high, as the reason already pointed out in Section \ref{Subsec:AnaUplinkCommRel}. In the following subsection, some numerical results and discussions will be provided to evaluate the performances of the downlink communication reliability for the PMCA scheme.

\subsection{Numerical Results and Discussions}\label{SubSec:NumResCommRel}

\begin{figure*}[!t]
	\centering
	\includegraphics[height=3.0in,width=\linewidth]{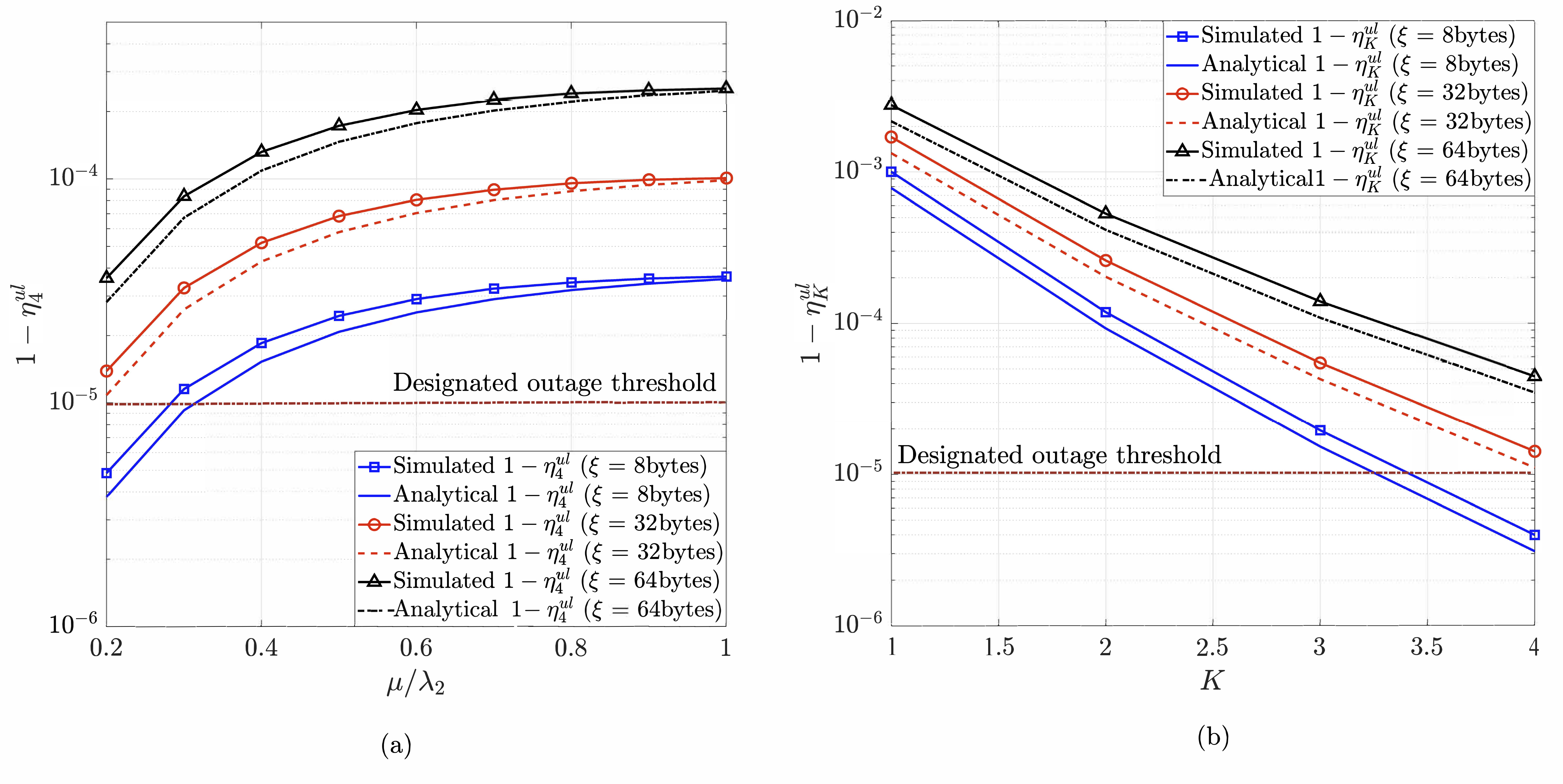}
	\caption{Numerical results of the uplink outage $1-\eta^{ul}_K$ for the case of non-collaborative APs: (a) $1-\eta^{ul}_K$ versus $\mu/\lambda_2$ for $K=4$, (b) $1-\eta^{ul}_K$ versus $K$ for $\mu/\lambda_2=0.5$ (users/small cell AP) and $\lambda_2=250$ (APs/km$^2$).}
	\label{Fig:UplinkRelNonCoop}
\end{figure*}

To validate the analytical results obtained in the previous subsections and evaluate the communication reliability performances of open-loop communication and PMCA for short packet transmission, some numerical results are provided in this subsection. The network parameters for simulation is shown in Table \ref{Tab:SimPara}.  To clearly show whether or not the target communication reliability of $99.999\%$ is attained in the uplink and downlink for different situations of PMCA, the following figures demonstrate the simulation results of the uplink and downlink outage probabilities (i.e., $1-\eta^{ul}_K$ and $1-\eta^{dl}_K$) and the designated outage threshold is thus $1-99.999\%=10^{-5}$. Since the simulation of the uplink and downlink outages is rare-event, it is terminated as the outage event occurs over 200 times so that we can obtain much confident statistics. Due to considering short packet transmission, we adopt the maximum  achievable rate of short blocklength regime without considering channel dispersion found in~\cite{CSCSCYTQYL19,WYGDTKYP14} to infer the outage condition for the HetNet in this paper as follows:
\begin{align}\label{Eqn:OutageShotPckTran}
 \ln(1+\mathrm{SIR}^{ul}_K)-\frac{Q^{-1}_G(\epsilon)}{\sqrt{\tau B}} \leq \frac{\xi}{B\tau},
\end{align}
where $\text{SIR}^{ul}_K$ equals $\max_{k\in\{1,\ldots,K\}}\{\gamma^{ul}_k\mathds{1}(V_k\in \mathcal{V}^{nc}_K)\}$ for uplink and $\max_{k\in\{1,\ldots,K\}}\{\gamma^{dl}_k\}$ for downlink, $Q^{-1}_G(\cdot)$ stands for the inverse of (Gaussian) $Q$-function, $\xi$ denotes the size of a short packet, and $\tau$ is the duration of transmission. The inequality in~\eqref{Eqn:OutageShotPckTran} can be further rewritten as
\begin{align}
\mathrm{SIR}^{ul}_K \leq \exp\left(\frac{Q^{-1}_G(\epsilon)}{\sqrt{\tau B}}+\frac{\xi}{B\tau}\right)-1.
\end{align}
Thus, we can get the uplink reliability of short packet transmission, i.e., $\eta^{ul}_K=\mathbb{P}[\mathrm{SIR}^{ul}_K\geq \theta]$ by setting the SIR threshold as $\theta = \exp\left(\frac{Q^{-1}_G(\epsilon)}{\sqrt{\tau B}}+\frac{\xi}{B\tau}\right)-1$, as shown in Table~\ref{Tab:SimPara}.

Figure~\ref{Fig:UplinkRelNonCoop} shows the numerical results of the uplink outage probabilities for the non-collaborative scenario. The analytical results corresponding to this scenario in the figure are found by using \eqref{Eqn:UplinkCommRelRel}. As shown in Fig.~\ref{Fig:UplinkRelNonCoop}(a), $1-\eta^{ul}_K$ increases ($\eta^{ul}_K$ decreases) as $\mu/\lambda_2$ increases. This is because more co-channel interference is created as $\mu/\lambda_2$ gets larger so that more APs are associated with users and become active. The short packet transmission with a longer packet size makes the outage probability increase, which can be mitigated by decreasing $\mu/\lambda_2$. Also, all the simulated results are very close to their corresponding analytical results, which validates that the analytical result in \eqref{Eqn:UplinkCommRelRel} is still fairly accurate even when the received uplink SIRs at different APs are assumed to be independent when deriving~\eqref{Eqn:UplinkCommRelRel}. Figure \ref{Fig:UplinkRelNonCoop}(b) illustrates how the uplink outage probability is suppressed as the number of APs in virtual cell increases. As shown in the figure, $1-\eta^{ul}_K$ significantly reduces as $K$ increases from 1 to 4, but only the case of the blue curves is able to achieve the outage probability below the designated outage threshold when $K\geq 4$. In light of this, users are suggested to associate with more APs and/or reduce their packet size.

\begin{figure*}[!t]
	\centering
	\includegraphics[height=3.0in, width=\linewidth]{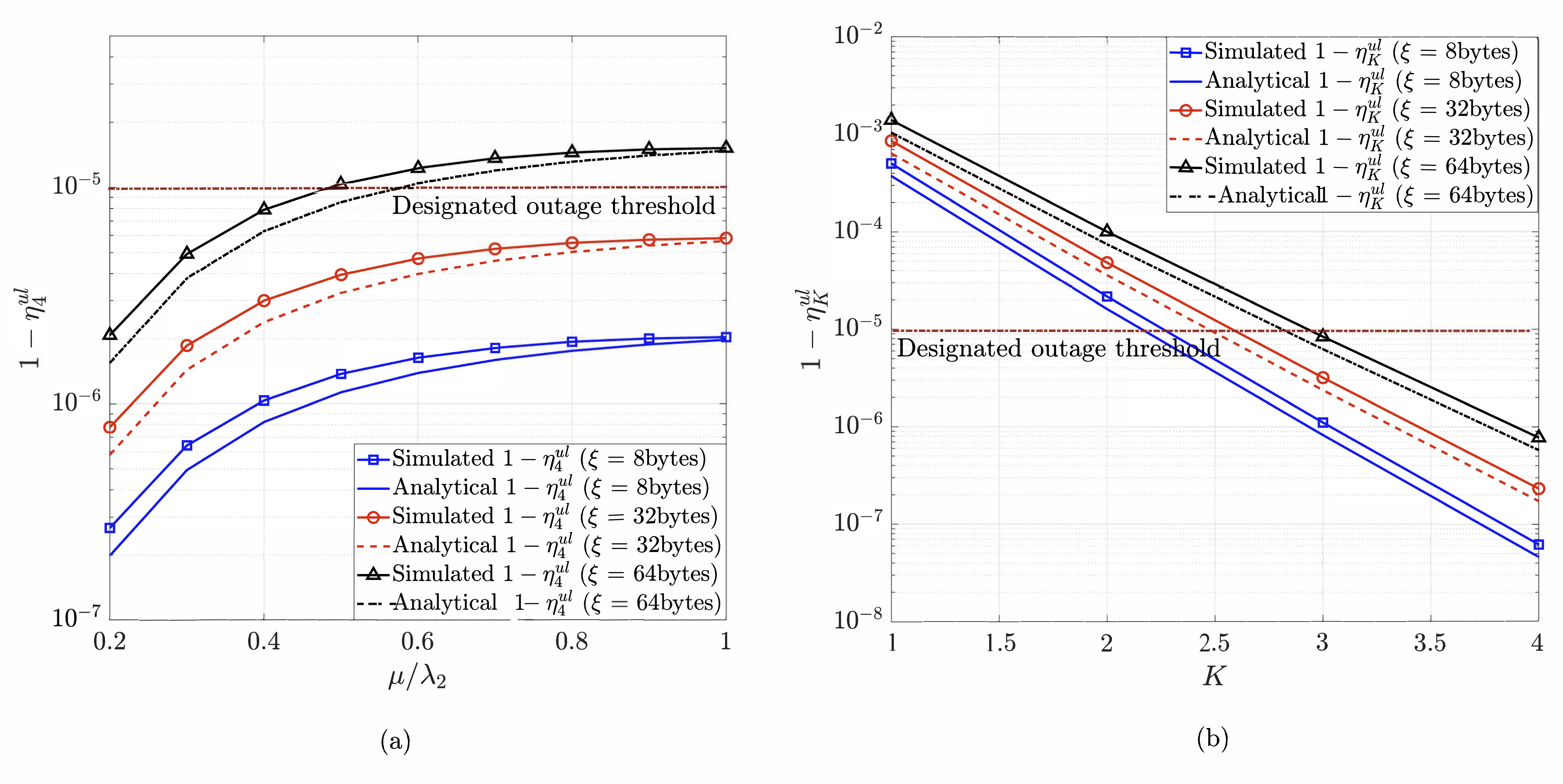}
	\caption{Numerical results of the uplink outage $1-\eta^{ul}_K$ for the scenario of collaborative APs: (a) $1-\eta^{ul}_K$ versus $\mu/\lambda_2$ for $K=4$, (b) $1-\eta^{ul}_K$ versus $K$ for $\mu/\lambda_2=0.5$ (users/small cell AP) and $\lambda_2=250$ (APs/km$^2$).}
	\label{Fig:UplinkRelCoop}
\end{figure*}

\begin{figure*}[!t]
	\centering
	\includegraphics[height=3.0in,width=\linewidth]{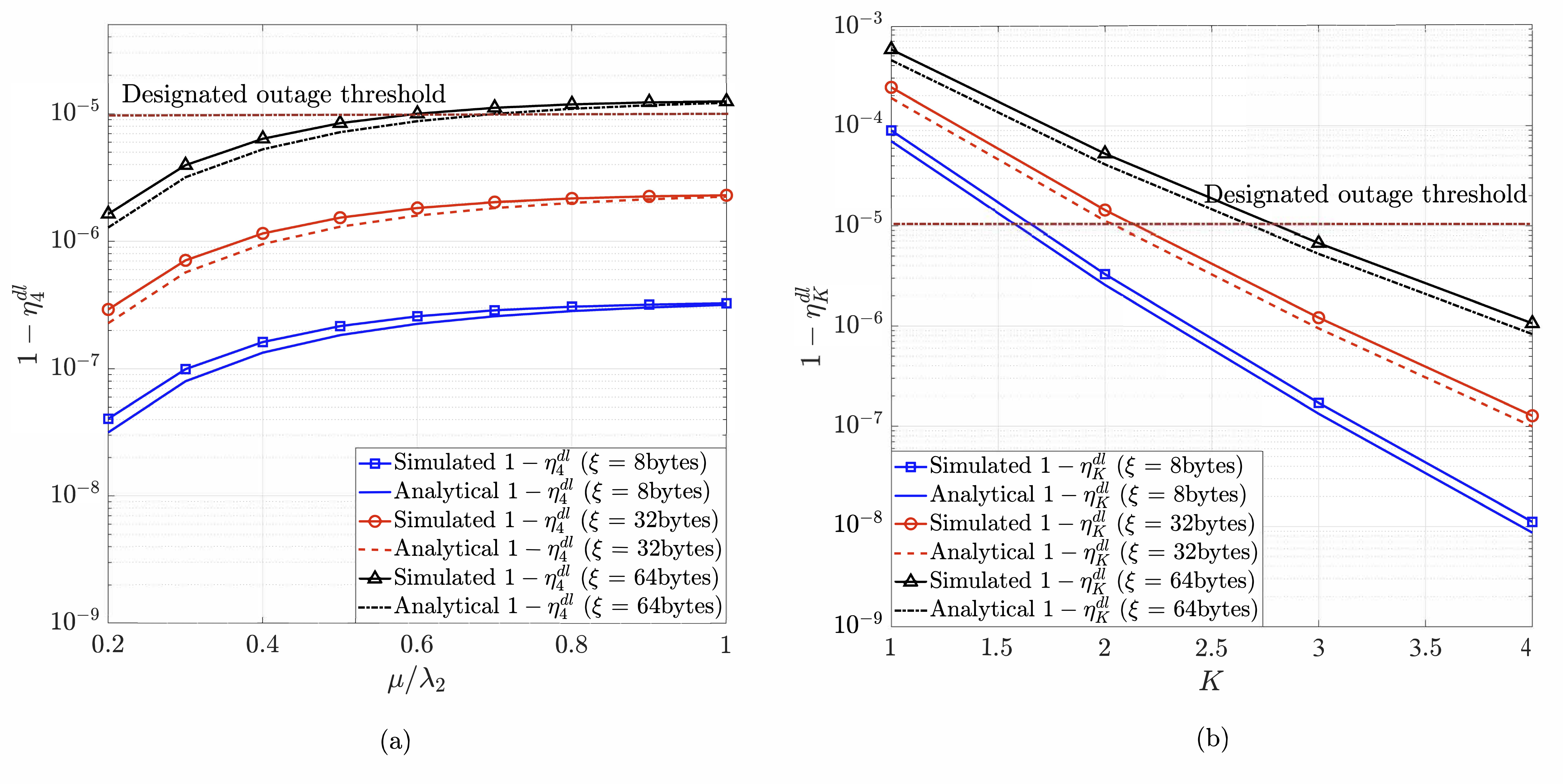}
	\caption{Numerical results of the downlink outage $1-\eta^{dl}_K$ for the scenario of non-collaborative APs: (a) $1-\eta^{dl}_K$ versus $\mu/\lambda_2$ for $K=4$, (b) $1-\eta^{dl}_K$ versus $K$ for $\mu/\lambda_2=0.5$ (users/small cell AP) and $\lambda_2=250$ (APs/km$^2$).}
	\label{Fig:DownlinkRelNonCoop}
\end{figure*}

\begin{figure*}[!t]
	\includegraphics[height=3.0in, width=\linewidth]{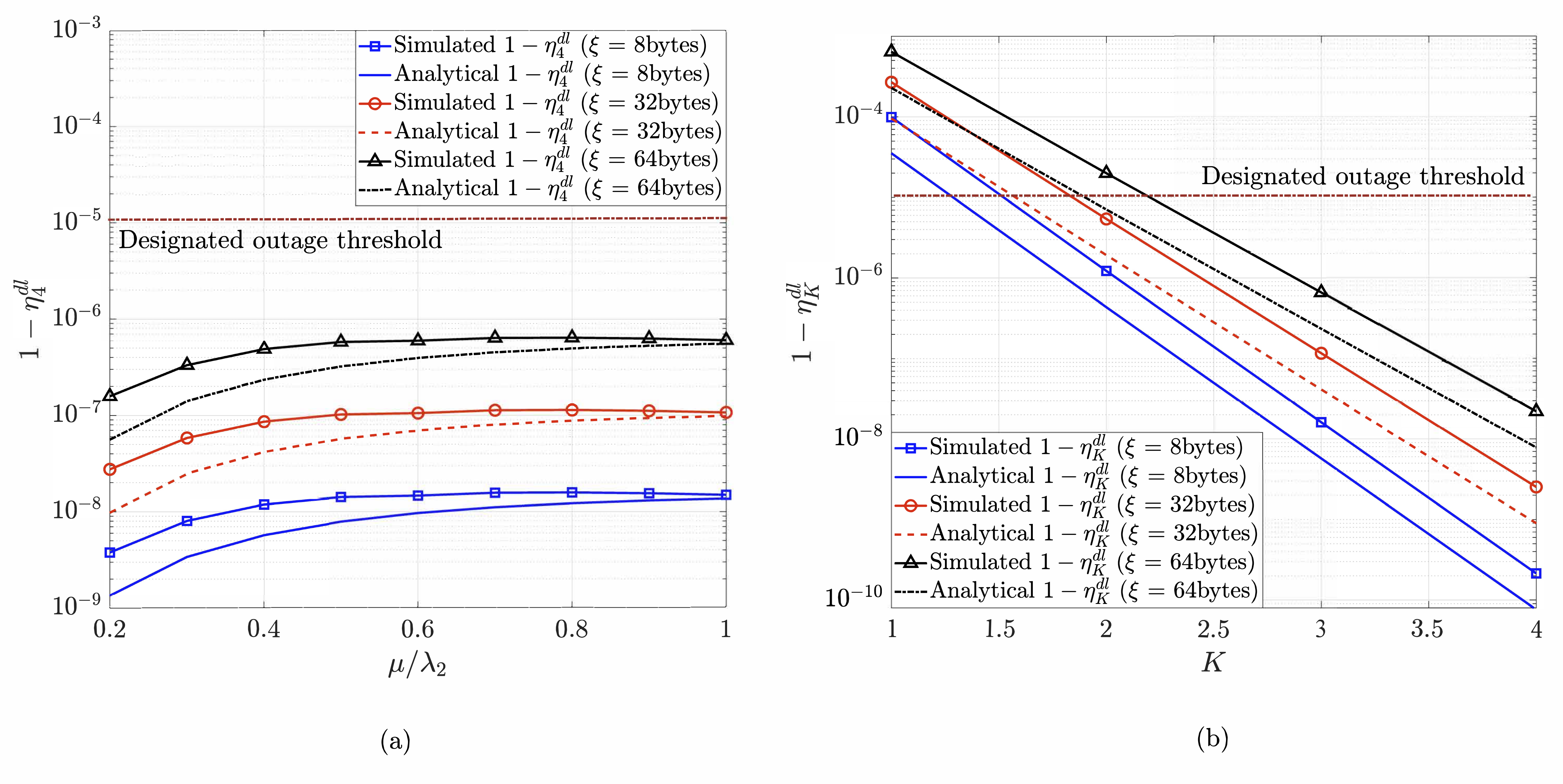}
	\caption{Numerical results of the downlink outage $1-\eta^{dl}_K$ for the scenario of collaborative APs: (a) $1-\eta^{dl}_K$ versus $\mu/\lambda_2$ for $K=4$, (b) $1-\eta^{dl}_K$ versus $K$ for $\mu/\lambda_2=0.5$ (users/small cell AP) and $\lambda_2=250$ (APs/km$^2$).}
	\label{Fig:DownlinkRelCoop}
\end{figure*}

\begin{figure*}[!t]
	\centering
	\includegraphics[height=3.0in, width=\linewidth]{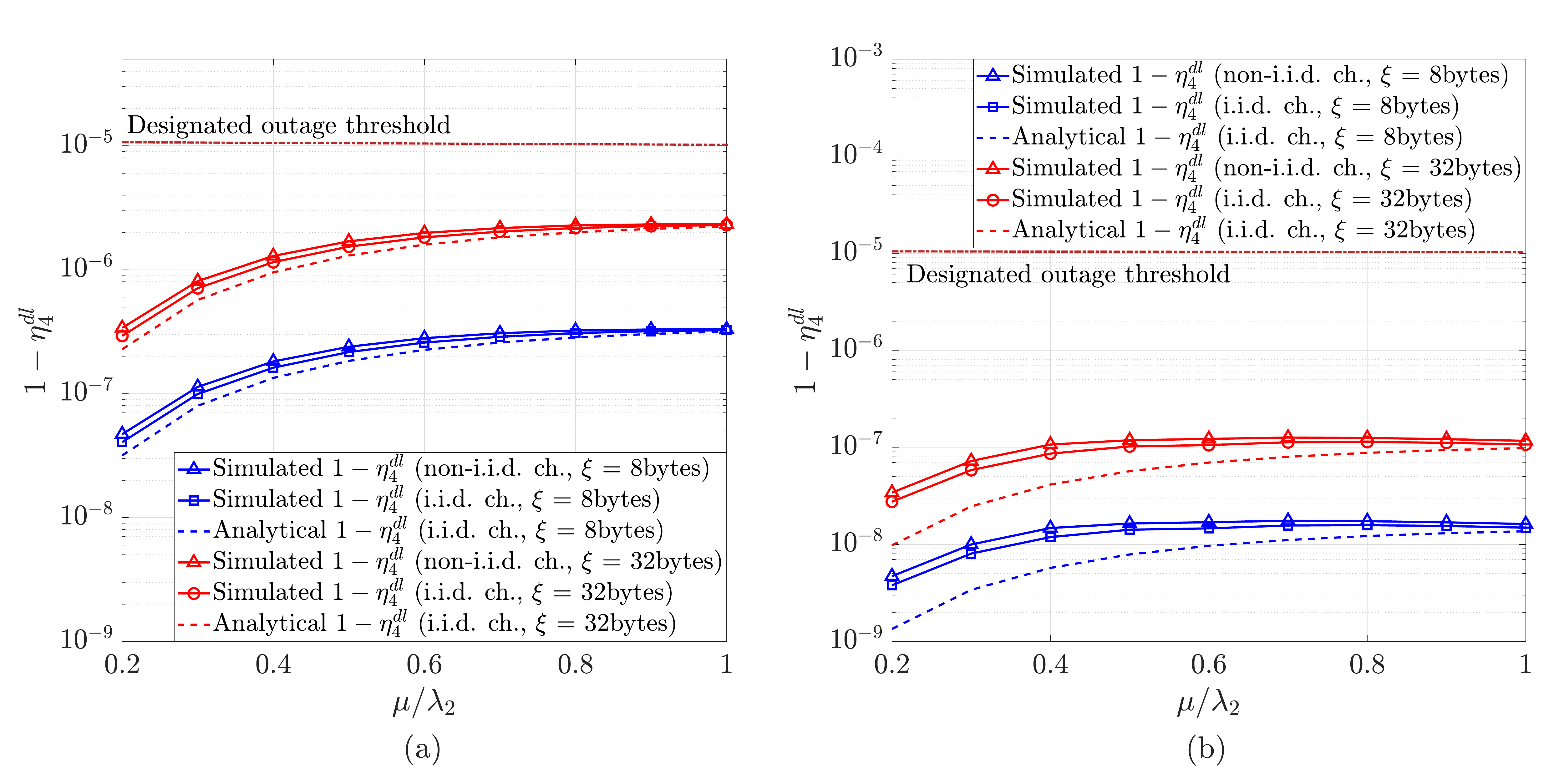}
	\caption{Numerical results of the downlink outage $1-\eta^{dl}_K$ when all the APs have spatially correlated fading channels: (a) $1-\eta^{dl}_K$ versus $\mu/\lambda_2$ for $K=4$ and non-collaborative APs, (b) $1-\eta^{dl}_K$ versus $\mu/\lambda_2$ for $K=4$ and collaborative APs.}
	\label{Fig:DownlinkOutageCorrChannels}
\end{figure*}

The simulated results of the uplink outage probabilities for the scenario of collaborative APs are shown in Fig. \ref{Fig:UplinkRelCoop} in which the analytical results that are the lower bound on the outage probability are calculated by using \eqref{Eqn:UplinkRelCop}. Only the results of the black curves in Fig.~\ref{Fig:UplinkRelCoop}(a) are above the designated outage threshold of $10^{-5}$ and thereby all the results in Fig.~\ref{Fig:UplinkRelCoop}(a) are much better than those in Fig.~\ref{Fig:UplinkRelNonCoop}(a). This reveals that we do not need to deploy many APs to significantly decrease $1-\eta^{ul}_K$ when the APs are able to collaborate. In Fig. \ref{Fig:UplinkRelCoop}(b), the simulated outcomes are also much better than those in Fig.~\ref{Fig:UplinkRelNonCoop}(b), which indicates that PMCA should be jointly implemented with CoMP so as to better serve the URLLC traffic. The simulated results of the downlink outage probabilities for the scenarios of non-collaborative and collaborative APs are shown in Figs. \ref{Fig:DownlinkRelNonCoop} and \ref{Fig:DownlinkRelCoop}, respectively. In general, they have similar ascending/descending curve trends if compared with the results in Figs. \ref{Fig:UplinkRelNonCoop} and \ref{Fig:UplinkRelCoop}, yet there are still some subtle discrepancies between them. The downlink outage probability, for instance, is lower than its uplink counterpart because users are not interfered by their first $K$ strongest APs owing to PMCA in the downlink, whereas APs are very likely to be severely interfered by their nearby users in the uplink. Also, APs can coordinate to allocate their RRUs in the downlink with the aid of anchor nodes so that their downlink channel collisions can be largely reduced. As such, the uplink outage is the main hurdle of achieving URLLC requirements by using open-loop communication and PMCA and we should use it to properly determine how densely the macro and small cell APs should be deployed in the HetNet. Moreover, another important implication that can be learned from Figs. 4(a)-7(a) is that all the uplink and downlink outage probabilities converge to constants as $\mu/\lambda_2\geq 1$, which represents a much practical scenario of the AP deployment that the density of users is smaller than that of small cell APs. Hence, the target communication reliability in the uplink and downlink can certainly be achieved in any practical situations of the AP deployment as long as the number of the collaborating APs in a virtual cell and the size of short packet transmission are properly chosen. Finally, we would like to illustrate whether or  not spatially correlated fading notably impacts the analyses of the communication reliability in Sections~\ref{Subsec:AnaUplinkCommRel} and~\ref{SubSec:AnaDLCommRel} that are conducted by assuming spatially independent fading between channels. We consider the distance-based correlated fading model. Namely, in the expression of the downlink SIR in~\eqref{Eqn:Downlink-kthSIR}, the fading channel gain $H_k$ is found as $H_k=\frac{H_1}{1+d_k^{\alpha}}+\frac{d_k^{\alpha}}{1+d_k^{\alpha}}H$ where $d_k$ is the distance between AP $V_k$ and AP $V_1$ and $H\sim\exp(1)$ is independent of all $H_k$'s and $H_j$'s\footnote{ The channel gain $H_j$ in interference $I^{dl}_K$ is also determined based on the same model as $H_k$. For the uplink case, we can also adopt a similar distance-based model to characterize the spatially correlated fading between uplink channels.}. The simulation results in Fig.~\ref{Fig:DownlinkOutageCorrChannels} (a) and (b) are corresponding to those in Fig.~\ref{Fig:DownlinkRelNonCoop}(a) and those in Fig.~\ref{Fig:DownlinkRelCoop}(a), respectively. As can be observed in Fig.~\ref{Fig:DownlinkOutageCorrChannels}, the outcomes with correlated fading are slightly worse than the analytical results with independent fading so that our analyses are still very accurate even though they do not consider spatially fading correlation between channels. This is because the spatially fading correlation between the channels of APs are usually fairly weak because most of APs are deployed far enough such that their channels are not very likely to fad at the same time.

\section{Modeling and Analysis of Communication Latency}\label{Sec:CommLatency}
The communication latency between an anchor node and a user is mostly contributed by the transmission delay between the anchor node and its associated APs and the communication delay between the APs and the user associated with them\footnote{Note that our focus in this section is to study the communication delays induced by the PMCA scheme and thus some delays not directly related to the PMCA scheme, such as signal processing delays on the transmitter and receiver sides are ignored.}.  The major latency difference between uplink and downlink is that uplink RRU collisions incur additional channel access delay. In the following, we will first develop a modeling and analysis approach to the uplink communication delay for the PMCA scheme, and then we apply a similar approach to characterize the downlink communication delay. 

\subsection{Uplink Communication Delay}
The uplink communication delay $D^{ul}$ mainly consists of the channel access delay $D^{ul}_{ac}$, uplink transmission delay $D^{ul}_{tr}$, and uplink backhaul delay $D^{ul}_{ba}$. It can be expressed as
\begin{align}
D^{ul} = D^{ul}_{ac}+D^{ul}_{tr}+D^{ul}_{ba}.
\end{align} 
Since there are $K$ APs in a virtual cell and a user has to randomly select an RRU for each of the $K$ APs, RRU collisions could happen in the cells of the $K$ APs. When a user starts to (randomly) select RRUs in its virtual cell, the channel access delay of the user can be defined as the lapse of time needed by the user to successfully access at least one non-collision RRU in its virtual cell. Recall that  $\rho^{ul}_K$ in \eqref{Eqn:NonColProb} is the uplink non-collision probability of the user. The mean of $D^{ul}_{ac}$ is equal to $1/\rho^{ul}_K$, i.e., $\mathbb{E}[D^{ul}_{ac}]= 1/\rho^{ul}_K$, which represents the average time for a user  to successfully access an RRU and then send a message. The uplink transmission delay $D^{ul}_{tr}$ is defined as the time duration between two messages that are successfully sent to at least one of the $K$ APs after the user successfully accesses at least one non-collision RRU in its virtual cell. We assume $D^{ul}_{tr}$ is ergodic so that the mean of $D^{ul}_{tr}$ can be found by $\mathbb{E}\left[D_{tr}^{ul}\right]=\lim_{T\rightarrow\infty}\frac{1}{T}\int_{0}^{T}D^{ul}_{tr}(t) \dif t $, i.e.,
\begin{align}
\mathbb{E}\left[D_{tr}^{ul}\right] &\defn \left[\lim_{T\rightarrow\infty}\frac{1}{T} \sum_{t=0}^{T}\mathds{1}\left(\max_{k}\{\gamma^{ul}_k(t)\}\geq \theta \right)\right]^{-1}\nonumber\\ 
&= \frac{1}{\mathbb{E}\left[\mathds{1}\left(\max_{k}\{\gamma^{ul}_k\}\geq \theta \right)\right]}=\frac{1}{\eta^{ul}_K}.
\end{align}
Note that the units of $\mathbb{E}\left[D_{ac}^{ul}\right]$ and $\mathbb{E}\left[D_{tr}^{ul}\right]$ can be properly transformed to seconds once the time duration (seconds) of transmitting a message is determined. 

The uplink backhaul delay is defined as the minimum transmission time needed for the APs in a virtual cell to transmit a message to their anchor node so that it can be mathematically expressed as
\begin{align}
D^{ul}_{ba} \defn \min_{k:V_k\in\mathcal{V}^{nc}_K}\{D^{ul}_{ba,k}\},
\end{align}
where $D^{ul}_{ba,k}$ denotes the uplink backhaul delay from the $k$th AP to its anchor node. We further assume that the arrival process of the messages from an AP to its anchor node can be modeled by an independent Poisson process, which gives rise to the fact that $D^{ul}_{ba,k}$ can be characterized by an exponential RV with some parameter $\beta$. In light of this, the distribution of $D^{ul}_{ba}$ in the non-collaborative AP case is
\begin{align*}
\mathbb{P}\left[D^{ul}_{ba}\leq x\right] &= 1- \mathbb{E}\left[\prod_{k=1}^{N} \mathbb{P}\left[D^{ul}_{ba,k}\geq x\right]\right]\\
&=1-\mathbb{E}\left[\exp\left(-x\beta N\right)\right]\\
&=1-\left(1-\eta^{ul}_K+\eta^{ul}_Ke^{-\beta x}\right)^K
\end{align*}
since all $D^{ul}_{ba,k}$'s are independent and $N$ that denotes the number of the non-collision APs  in the virtual cell is a binomial RV with parameters $K$ and $\eta^{ul}_K$. We thus have the mean of $D^{ul}_{ba}$ found as follows:
\begin{align}
\mathbb{E}[D^{ul}_{ba}] &=\int_{0}^{\infty}\mathbb{P}\left[D^{ul}_{ba}\geq x\right]\dif x \nonumber\\
&=\int_{0}^{\infty} \left(1-\eta^{ul}_K+\eta^{ul}_Ke^{-\beta x}\right)^K\dif x.
\end{align}
If $\eta^{ul}_K\approx 1$, we can get $\mathbb{E}[D^{ul}_{ba}]\approx 1/\beta K$.   For the case of coordinated APs, $\mathbb{P}\left[D^{ul}_{ba}\geq x\right] =\exp(-x\beta )$ because only one message is sent from the virtual cell to the anchor node. The mean of $D^{ul}_{ba}$ is $\mathbb{E}[D^{ul}_{ba}]=1/\beta$ and the unit of $\mathbb{E}[D^{ul}_{ba}]$ can be set as times/RRU. The mean of the uplink communication delay is readily approximated as
\begin{align}
\mathbb{E}\left[D^{ul}\right] \approx
&\begin{cases}
\int_{0}^{\infty} \left(1-\eta^{ul}_K+\eta^{ul}_Ke^{-\beta x}\right)^K\dif x, &\text{Non-collab.}\\
\frac{1}{\beta}, &\text{Collab.}
\end{cases},\nonumber\\
& +\frac{1}{\rho^{ul}_K}+\frac{1}{\eta^{ul}_K},
\end{align}
which can be employed to evaluate the uplink latency performance. From the above result, we can see that the mean uplink communication latency is inversely proportional to the uplink communication reliability, which is also shown in~\cite{XG19} by using the queuing theory in an autonomous vehicular network.   

\begin{figure*}[!t]
	\centering
	\includegraphics[height=2.75in,width=\linewidth]{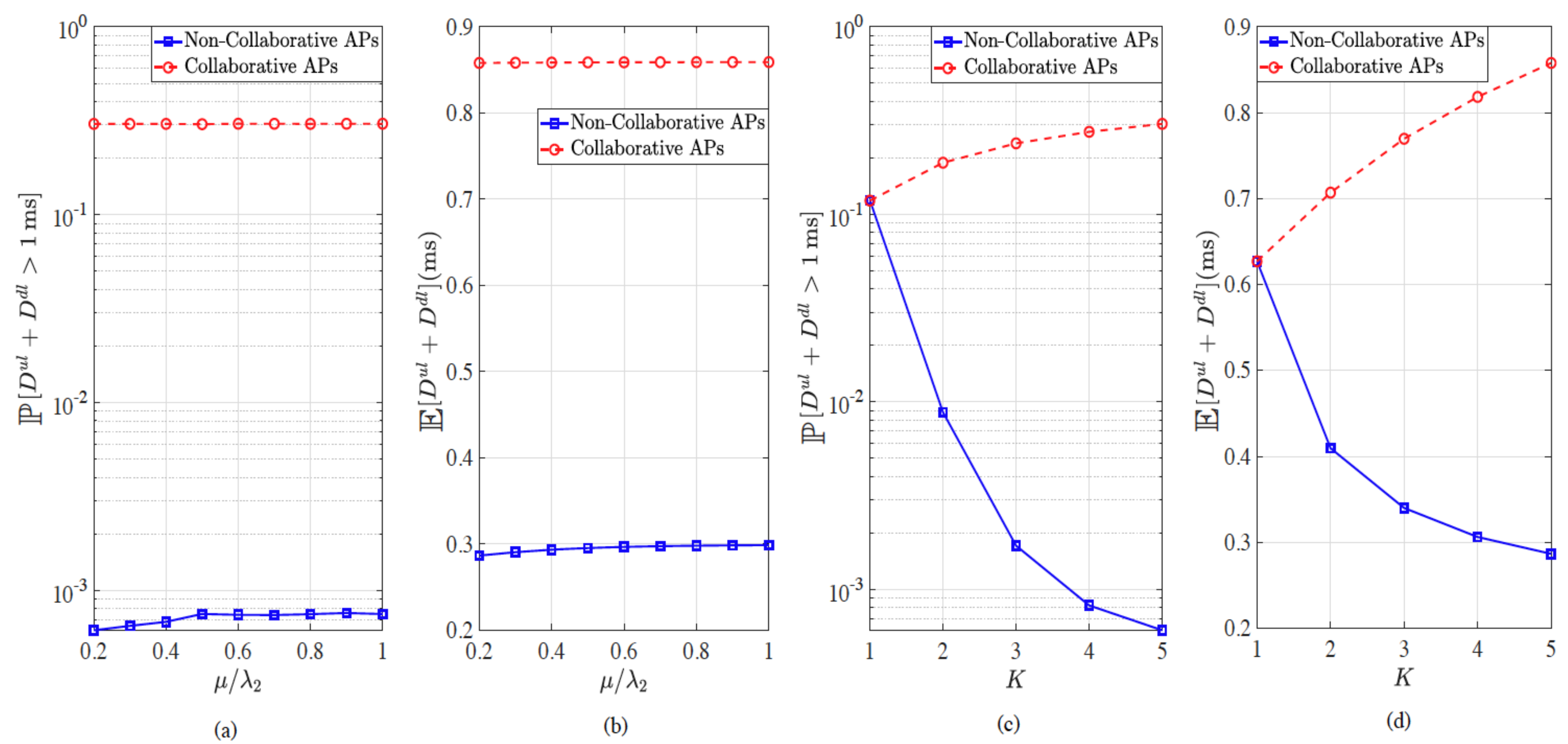}
	\caption{Numerical results of $D^{ul} + D^{dl}$ for showing the latency performances of collaborative APs and non-collaborative APs: (a) $\mathbb{P}[D^{ul}+D^{dl} > 1\text{ ms}]$ versus $\mu/\lambda_2$ for $K=5$, (b) $\mathbb{E}[D^{ul}+D^{dl}]$ versus $\mu/\lambda_2$ for $K=5$, (c) $\mathbb{P}[D^{ul}+D^{dl} > 1\text{ ms}]$ versus $K$ for $\mu/\lambda_2=0.2$ (users/small cell AP) and $\lambda_2=250$ (APs/km$^2$), (d) $\mathbb{E}[D^{ul}+D^{dl}]$ versus $K$ for $\mu/\lambda_2=0.2$ (users/small cell AP) and $\lambda_2=250$ (APs/km$^2$).}
	\label{Fig:MeanDelays}
\end{figure*}

\subsection{Downlink Communication Delay}
In the downlink, since each AP is able to allocate its resource to its received message, the downlink communication delay $D^{dl}$ mostly consists of the downlink backhaul delay $D^{dl}_{ba}$ and transmission delay $D^{dl}_{tr}$, i.e., it can be simply written as
\begin{align}
D^{dl} = D^{dl}_{ba}+D^{dl}_{tr}.
\end{align}
The downlink backhaul delay is defined as the maximum time elapsed from the start time of sending a message from the anchor node to the end time when all $K$ APs in the virtual cell receive the message. Suppose the message arrival process at each AP can be modeled as an independent Poisson process and the downlink backhaul delay can be expressed as
\begin{align}
D^{dl}_{ba}\defn 
\begin{cases}
\max_{k\in\{1,\ldots,K\}}\{D^{dl}_{ba,k}\},&\text{Collab. APs}\\
\min_{k\in\{1,\ldots,K\}}\{D^{dl}_{ba,k}\},&\text{Non-collab. APs},
\end{cases}
\end{align}
where $D^{dl}_{ba,k}\sim\exp(\beta)$ is the downlink backhaul delay from the anchor node to the $k$th AP in the virtual cell. In light of this, the distribution of $D^{dl}_{ba}$ is found as
\begin{align*}
\mathbb{P}[D^{dl}_{ba}\geq x] &=
\begin{cases}
1-\mathbb{P}\left[\max_{k\in\{1,\ldots,K\}}\{D^{dl}_{ba,k}\}\leq x\right], \\
\mathbb{P}\left[\min_{k\in\{1,\ldots,K\}}\{D^{dl}_{ba,k}\}\geq x\right], 
\end{cases}\\
 &=\begin{cases}
 1-(1-e^{-\beta x})^K, \\
e^{-\beta K x}
 \end{cases}
\end{align*}
The downlink transmission delay is the time duration in which the $K$ APs in the virtual cell successfully transmit a message to the user and its mean can be characterized by the downlink communication reliability, i.e., $\mathbb{E}[D^{dl}_{tr}]=1/\eta^{dl}_K$ and its bound can be found by using Theorem \ref{Thm:DownlinkReliability}. Accordingly, the mean of the downlink communication delay is given by
\begin{align}
\mathbb{E}\left[D^{dl}\right]=&
\begin{cases}
\int_{0}^{\infty} \left[1-\left(1-e^{-\beta x}\right)^K\right] \dif x, &\text{Collab.APs}\\
\frac{1}{\beta K}, &\text{Non-collab. APs}
\end{cases}\nonumber\\
&+\frac{1}{\eta^{dl}_K}.
\end{align} 
Hence, $\mathbb{E}\left[D^{dl}\right]$ increases as $K$ increases in the case of collaborative APs, but it decreases as $K$ increases in the case of non-collaborative APs. 

\subsection{Numerical Results}

In this subsection, we would like to numerically demonstrate how the statistical properties of $D^{ul}+D^{dl}$ vary with $\mu/\lambda_2$ and $K$. The network parameters in Table \ref{Tab:SimPara}, URLLC short packets with 32 bytes, and $\beta = 5$ messages/ms are adopted for simulation. Figure \ref{Fig:MeanDelays} shows the numerical results of delay outage probability $\mathbb{P}[D^{ul}+D^{dl} >1\text{ ms}]$ and mean delay $\mathbb{E}[D^{ul}+D^{dl}]$. As can be seen in Figs. \ref{Fig:MeanDelays}(a) and \ref{Fig:MeanDelays}(b), the delay outage probability and mean delay both slightly vary as $\mu/\lambda_2$ changes and obviously the latency performance in the case of non-collaborative APs outperforms that in the case of collaborative APs. Thus, the latency performance of the PMCA scheme is mainly dominated by whether or not the APs in a virtual cell can collaborate since more networking delay is incurred due to AP coordination. This phenomenon can be clearly observed in the numerical results in Figs.~\ref{Fig:MeanDelays}(c) and \ref{Fig:MeanDelays}(d). As shown in Figs.~\ref{Fig:MeanDelays}(c) and \ref{Fig:MeanDelays}(d), the delay outage probability and the mean delay in the case of collaborative APs largely degrade as $K$ increases if compared with those in the case of non-collaborative APs. Therefore, we can draw a conclusion that there exists a trade-off problem between the communication reliability and latency, which needs to be aware when the PMCA scheme is employed in practice.

\section{Conclusions}\label{Sec:Conclusion}
In this paper, we break the longstanding concept that a tradeoff between communication reliability and latency always exists the in cellular systems and shed the light on the fact that extremely reliable communication is hardly benefited by receiver feedback. Our main contribution is to first claim that ultra-reliable open-loop communication is the key to ultimately fulfilling the goal of ultra-low latency in the network and then devise an analytical framework to validate the claim. The PMCA scheme and corresponding open-loop communication protocols in a HetNet are proposed and how much the communication reliability and latency of a virtual cell achieved by them is analyzed. Our analytical outcomes and numerical results show that the proposed open-loop communication and PMAC scheme are able to achieve the target reliability and latency of URLLC users in a HetNet provided that the APs are sufficiently deployed and the number of the APs in a virtual cell is properly chosen.

\appendix [Proofs of Lemmas and Theorems]

\subsection{Proof of Lemma \ref{Lem:PMFUserAssKAPs}}\label{App:PMFUserAssKAPs}
According to Theorem 1 in \cite{CHLKLF16}, we can obtain the following result:
\begin{align*}
&\mathbb{P}\left[\max_{m,i:A_{m,i}\in\Phi} w_m\|A_{m,i}\|^{-\alpha}\leq x^{-\alpha}\right]\\
&=\mathbb{P}\left[\min_{m,i:A_{m,i}\in\Phi} w^{\frac{1}{\alpha}}_m\|A_{m,i}\|\geq x \right]\\
&=\mathbb{P}\left[\|\widetilde{A}_1\|\geq x\right]= \exp\left(-\pi x^2\sum_{m=1}^{2}w^{\frac{2}{\alpha}}_m\lambda_m\right),
\end{align*}
where $\|\widetilde{A}_1\|\defn \min_{m,i:A_{m,i}\in\Phi} w^{\frac{1}{\alpha}}_m\|A_{m,i}\|$ and $\|\widetilde{A}_1\|^2\sim\exp(\pi\sum_{m=1}^{2}w^{2/\alpha}_m\lambda_m)$ is an exponential random variable with parameter $\pi\sum_{m=1}^{2}w^{2/\alpha}_m\lambda_m$. Let $\widetilde{\Phi}\defn\{\widetilde{A}_k\in\mathbb{R}^2: k\in\mathbb{N}_+\}$ be a homogeneous PPP of density $\sum_{m=1}^{2}w^{2/\alpha}_m\lambda_m$ and $\widetilde{A}_k$ is the $k$th nearest point in $\widetilde{\Phi}$ to the origin. Also, we define $\widetilde{\Phi}_m\defn\{\widetilde{A}_{m,k}\in\mathbb{R}^2: \widetilde{A}_{m,k}=w^{1/\alpha}_m\widetilde{A}_k, \widetilde{A}_k\in\widetilde{\Phi}, k\in\mathbb{N}_+\}$ and it is a homogeneous PPP of density $\widetilde{\lambda}_m\defn\sum_{i=1}^{2}(w_i/w_m)^{2/\alpha}\lambda_i$ based on the result of Theorem 1 in \cite{CHLLCW16}. This means that $\widetilde{\Phi}_m$ can be viewed as a sole homogeneous PPP equivalent to the superposition of all independent homogeneous PPPs in the network (i.e., $\bigcup_{i=1}^2\Phi_i$) when all tier-$i$ APs in $\Phi_i$ are  scaled by $(w_m/w_i)^{2/\alpha}$. Thus, \eqref{Eqn:kthStrongestAP} can be equivalently expressed as $\|V_k\| \stackrel{d}{=} \|\widetilde{A}_{m,k}\|,\,\,k=1,2,\ldots,$
where $\stackrel{d}{=}$ stands for the equivalence in distribution. This result also indicates that the typical user can be imaged to equivalently associate with the first $K$ nearest APs in $\widetilde{\Phi}_m$. 

Since the typical user can associate with its first $K$ weighed nearest APs in the network, the distance between the typical user and its $K$th weighed nearest AP is $\|V_K\|$ and $\|V_K\|^2\stackrel{d}{=}\|\widetilde{A}_{m,K}\|^2$ where $\|\widetilde{A}_{m,K}\|^2$ is the sum of $K$ i.i.d. exponential random variables (RVs) which have the same distribution as $\|\widetilde{A}_{m,1}\|^2$ and thus $\|V_K\|^2\sim\text{Gamma}(K,\pi\widetilde{\lambda}_m)$ is a Gamma RV with shape $K$ and rate $\pi\widetilde{\lambda}_m$. Let $\mathcal{C}_{m,i}$ denote the cell area of AP $\widetilde{A}_{m,i}$ in which all users associate with $\widetilde{A}_{m,i}$ when the PMCA scheme is adopted and each user associates with its first $K$ nearest APs in $\widetilde{\Phi}_m$.  From \cite{CHLLCW16} and \cite{JSFNZ07}, we learn that the Lebesgue measure of $\mathcal{C}_{m,i}$, denoted by $\nu(\mathcal{C}_{m,i})$, for $K=1$ can be accurately described by a Gamma RV with the following pdf for all $i\in\mathbb{N}_+$:
\begin{align*}
f_{\nu(\mathcal{C}_{m})}(x)\approx\frac{(\zeta \widetilde{\lambda}_m x)^{\zeta}}{x\Gamma(\zeta)}e^{-\zeta x\widetilde{\lambda}_m},\quad\text{for } K=1,
\end{align*}
where $\zeta=\frac{7}{2}$. Since the mean of $\nu(\mathcal{C}_{m,i})$ for $K=1$ is $1/\widetilde{\lambda}_m$ and it is also equal to $\mathbb{E}[\pi \|V_1\|^2]$, the mean of $\nu(\mathcal{C}_{m,i})$ for $K>1$ is equal to $\mathbb{E}[\pi \|V_K\|^2]=K/\widetilde{\lambda}_m$. Accordingly, for $K>1$, $f_{\nu(\mathcal{C}_{m})}$ must be equal to
\begin{align*}
f_{\nu(\mathcal{C}_{m})}(x)\approx\frac{(\zeta_m x\widetilde{\lambda}_m/K )^{\zeta_m}}{x\Gamma(\zeta_m)}e^{-\zeta_m x\widetilde{\lambda}_m/K},\quad\text{for } K>1.
\end{align*}
Note that All $\nu(\mathcal{C}_{m,i})$'s have the same distribution and they may not be independent. Let $\widetilde{\Phi}_m(\mathcal{C}_{m,i})$ denote the number of users associating with $\widetilde{A}_{m,i}$ so that $p_{m,n}$ for $K>1$ can be expressed as
\begin{align*}
p_{m,n} &\defn \mathbb{P}[N_m=n] =\mathbb{P}[\widetilde{\Phi}_m(\mathcal{C}_{m,i})=n]\\
&=\mathbb{E}\left[\frac{(\widetilde{\lambda}_m\nu(\mathcal{C}_m)\mu)^n}{n!}\exp(-\widetilde{\lambda}_m\nu(\mathcal{C}_m)\mu)\right],
\end{align*}
which can be completely carried out by using the above expression of $f_{\nu(\mathcal{C}_{m})}(x)$ for $K>1$. Thus, the result in \eqref{Eqn:PMFUserAssKAPs} is obtained.

\subsection{Proof of Theorem \ref{Thm:LapTramsSK}}\label{App:ProofLapTransSK}
(i) According to the definition of $S_{-K}$, $S_{-K}$ can be alternatively written as
\begin{align*}
S_{-K} &=\sum_{i=K+1}^{\infty} H_iW_i\|V_i\|^{-\alpha}\\
&\stackrel{d}{=} \sum_{j:\widetilde{V}_j\in\widetilde{\Phi}} H_j(\|\widetilde{V}_K\|^2+\|\widetilde{V}_{j}\|^2)^{-\frac{\alpha}{2}},
\end{align*}
where $\stackrel{d}{=}$ denotes the equivalence in distribution, $\widetilde{\Phi}\defn\{\widetilde{V}_j\in\mathbb{R}^2: \widetilde{V}_j=W_jV_j, V_j\in\Phi, j\in\mathbb{N}_+\}$ is a homogeneous PPP of density $\widetilde{\lambda}$, and $\|\widetilde{V}_{j+i}\|^2=\|\widetilde{V}_j\|^2+\|\widetilde{V}_i\|^2$ for all $i, j\in\mathbb{N}_+$ and $i\neq j$ based on the proof of Proposition 1 in \cite{CHLDCL18}. Therefore, the Laplace transform of $S_{-K}$ can be calculated as shown in the following:
\begin{align*}
&\mathcal{L}_{S_{-K}}(s) =\mathbb{E}\left[\exp\left(\frac{-s}{\|\widetilde{V}_K\|^{\alpha}} \sum_{j:\widetilde{V}_j\in\widetilde{\Phi}} H_j \left(1+\frac{\|\widetilde{V}_j\|^2}{\|\widetilde{V}_K\|^2}\right)^{-\frac{\alpha}{2}}\right)\right]\\
&\stackrel{(a)}{=}\mathbb{E}_{Y_K}\left[\exp\left(-\pi\widetilde{\lambda}\int_{0}^{\infty}\mathbb{E}_{H}\left[1-e^{HsY_K^{-\frac{\alpha}{2}}(1+\frac{r}{Y_K})^{-\frac{\alpha}{2}}}\right]\dif r\right)\right]\\
&\stackrel{(b)}{=} \mathbb{E}_{Y_K}\left[\exp\left(-\pi\widetilde{\lambda}Y_K\int_1^{\infty}\mathbb{P}\left[Y_Kr'\leq \left(\frac{s H}{Z}\right)^{\frac{2}{\alpha}}\right] \dif r'\right)\right]\\
&\stackrel{(c)}{=}\mathbb{E}_{Y_K}\left\{\exp\left[-\pi\widetilde{\lambda} Y_K\ell\left(\frac{s}{Y_K^{\frac{\alpha}{2}}} ,\frac{2}{\alpha}\right)\right]\right\},
\end{align*}
where $\stackrel{(a)}{=}$ follows from the probability generating functional (PGFL) of the homogeneous PPP $\widetilde{\Phi}$\cite{MH12}\cite{FBBBL10} and $Y_k\defn \|\widetilde{V}_k\|^2$, $(b)$ is obtained by using $Z\sim\exp(1)$, and $(c)$ is obtained by using the derivation technique in the proof of Proposition 2 in \cite{CHLLCW16}. Thus, $\mathcal{L}_{S_{-K}}(s)$ is equal to the result in \eqref{Eqn:LapTransS-K} because $Y_k\sim\text{Gamma}(k,\pi\widetilde{\lambda})$.\\
(ii) Next, we find the Laplace transform of $S_K$. According to the definition of $V_k$ in \eqref{Eqn:kthStrongestAP}, we know that $S_{\infty}\defn \lim_{K\rightarrow\infty}S_K$ in \eqref{Eqn:TrunKthShotNoiseProc} can be equivalently expressed as
\begin{align*}
S_{\infty} \stackrel{d}{=}\sum_{m,i:A_{m,i}\in\Phi}H_{m,i}w_m\|A_{m,i}\|^{-\alpha},
\end{align*}
where all $H_{m,i}$'s are i.i.d. RVs with the same distribution as $H_k$ and $\Phi\defn \bigcup_{m=1}^2\Phi_m$. Thus, $\mathcal{L}_{S_{\infty}}(s)$ can be found as follows:
\begin{align*}
\mathcal{L}_{S_{\infty}}(s) &= \mathbb{E}\left[\exp\left(-s\sum_{m,i:A_{m,i}\in\Phi}H_{m,i}w_m\|A_{m,i}\|^{-\alpha}\right)\right]\\
&\stackrel{(a)}{=}\mathbb{E}\left[\exp\left(-s\sum_{k:\widetilde{V}_k\in\widetilde{\Phi}}H_k\|\widetilde{V}_k\|^{-\alpha}\right)\right]\\
&\stackrel{(b)}{=} \exp\left[-\pi\widetilde{\lambda}s^{\frac{2}{\alpha}}\mathbb{E}\left[H^{\frac{2}{\alpha}}\right]\Gamma\left(1-\frac{2}{\alpha}\right)\right]\\
&=\exp\left[-\frac{\pi\widetilde{\lambda}s^{\frac{2}{\alpha}}}{\mathrm{sinc}(2/\alpha)}\right],
\end{align*}
where $(a)$ follows from Theorem 1 \cite{CHLKLF16} and (b) follows from the PGFL of the homogeneous PPP $\widetilde{\Phi}$. Next, $\mathcal{L}_{S_{-K}}(s)$ can be derived as shown in the following: 
\begin{align*}
\mathcal{L}_{S_{-K}}(s) &=\mathbb{E}\left[\exp\left(-s\sum_{k:V_k\in\mathcal{V}_{\infty}\setminus\mathcal{V}_K} \frac{H_kW_k}{\|V_k\|^{\alpha}}\right)\right]\\
&=\mathbb{E}\left[\exp\left(-s\sum_{k:\widetilde{V}_k\in\widetilde{\mathcal{V}}_{\infty}\setminus\widetilde{\mathcal{V}}_K} \frac{H_k}{\|\widetilde{V}_k\|^{\alpha}}\right)\right],
\end{align*}
where $\widetilde{\mathcal{V}}_{\infty}\defn\{\widetilde{V}_k: k\in\mathbb{N}_+\}$ is a homogeneous PPP of density $\widetilde{\lambda}$, $\widetilde{V}_k$ is the $k$th nearest point in $\widetilde{\mathcal{V}}_{\infty}$ to the typical user and $\widetilde{\mathcal{V}}_K\defn\{\widetilde{V}_1,\ldots,\widetilde{V}_K\}$. Since  $\|\widetilde{V}_k\|^2\sim\text{Gamma}(k,\pi\widetilde{\lambda})$ is the sum of $k$ i.i.d. exponential RVs with parameter $\pi\widetilde{\lambda}$, we can get
\begin{align*}
&\mathbb{E}\left[\exp\left(-s\sum_{k:\widetilde{V}_k\in\widetilde{\mathcal{V}}_{\infty}\setminus\widetilde{\mathcal{V}}_K}  \frac{H_k}{\|\widetilde{V}_k\|^{\alpha}}\right)\right]\\
=&\mathbb{E}\left[\exp\left(-\frac{s}{\|\widetilde{V}_{K}\|^{\alpha}}\sum_{k:\widetilde{V}_k\in\widetilde{\mathcal{V}}_{\infty}} \frac{H_k}{(1+\|\widetilde{V}_k\|^2/\|\widetilde{V}_{K}\|^2)^{\frac{\alpha}{2}}}\right)\right]\\
=&\int_0^{\infty} \mathbb{E}\left[\prod_{k:\widetilde{V}_k\in\widetilde{\mathcal{V}}_{\infty}}\exp\left(- \frac{sx^{-\frac{\alpha}{2}}H_k}{(1+\frac{\|\widetilde{V}_k\|^2}{x})^{\frac{\alpha}{2}}}\right)\right]f_{\|\widetilde{V}_K\|^2}(x)\dif x \\
\stackrel{(c)}{=}& \mathbb{E}_{Y_K}\left\{e^{-\pi\widetilde{\lambda}Y_K\ell\left(sY_K^{-\frac{\alpha}{2}},\frac{2}{\alpha}\right)}\right\}
\end{align*}
where $(c)$ is obtained by first finding the PGFL of $\widetilde{\mathcal{V}}_{\infty}$ and we then follow the derivation steps in the proof of Proposition 4 in \cite{CHLDCL18} to derive function $\ell(\cdot,\cdot)$.  
In addition,  we can know the following:
\begin{align*}
\mathcal{L}_{S_{\infty}}(s) =&\mathbb{E}\left[e^{-s(S_K+S_{-K})}\right]\\
=&\mathbb{E}\left[\exp\left(-sS_K\right)\cdot \exp\left(-s\sum_{k:\widetilde{V}_k\in\widetilde{\mathcal{V}}_{\infty}\setminus\widetilde{\mathcal{V}}_K} \frac{H_k}{\|\widetilde{V}_k\|^{\alpha}}\right)\right]\\
=&\mathbb{E}_{\|\widetilde{V}_K\|^2}\bigg\{\exp\left(-sS_K\right)\\
&\times\mathbb{E}\left[e^{-s\sum_{k:\widetilde{V}_k\in\widetilde{\mathcal{V}}_{\infty}\setminus\widetilde{\mathcal{V}}_K} H_k\|\widetilde{V}_k\|^{-\alpha}}\bigg|\widetilde{V}_K\right]\bigg\}\\
=& \mathbb{E}_{Y_K^2}\left\{\exp\left[-sS_K-\pi\widetilde{\lambda}Y_K\ell\left(sY_K^{-\frac{\alpha}{2}},\frac{2}{\alpha}\right)\right]\right\}\\
=&\exp\left[-\frac{\pi\widetilde{\lambda}s^{\frac{2}{\alpha}}}{\text{sinc}(2/\alpha)}\right],
\end{align*}
which yields 
\begin{align*}
\mathcal{L}_{S_K}(s) =& \exp\left[-\frac{\pi\widetilde{\lambda}s^{\frac{2}{\alpha}}}{\text{sinc}(2/\alpha)}\right] \\
 &\times \mathbb{E}\left\{\exp\left[\pi\widetilde{\lambda}Y_K\ell\left(sY_K^{-\frac{\alpha}{2}},\frac{2}{\alpha}\right)\right]\right\}
\end{align*},
and it can be expressed as \eqref{Eqn:LapTransSK} due to $Y_K\sim\text{Gamma}(K,\pi\widetilde{\lambda})$.\\
(iii) For the upper bound on $\mathbb{P}[S_K\geq y]$ for $y\geq 0$, we can find it by using the following inequality:
\begin{align*}
\mathbb{P}\left[\sum_{k=1}^{K}Z_k\geq y\right] & =1- \mathbb{P}\left[\sum_{k=1}^{K}Z_k\leq y\right]\\
&\leq 1- \prod_{k=1}^{K}\mathbb{P}\left[Z_k \leq z_ky\right],
\end{align*}
where all $Z_k$'s are non-negative RVs, $z_k\in[0,1]$ for all $k\in\{1,2,\ldots,K\}$ and $\sum_{k=1}^{K}z_k=1$. By using the above inequality, the upper bound on $\mathbb{P}[S_K\geq y]$ can be thereupon found as follows:
\begin{align*}
\mathbb{P}[S_K\geq y]&\leq  1-\prod_{k=1}^{K}\mathbb{P}\left[H_kW_k\|V_k\|^{-\alpha}\leq z_ky\right]\\
&= 1-\prod_{k=1}^{K}\mathbb{P}\left[H_k\leq z_ky\|\widetilde{V}_k\|^{\alpha}\right]\\
&=1-\prod_{k=1}^{K}\mathbb{E}\left[1-\exp\left(-yz_kY^{\frac{\alpha}{2}}_k\right)\right]\\
&\stackrel{(d)}{=}1-\prod_{k=1}^{K}\left[1-\mathcal{L}_{Y^{\frac{\alpha}{2}}_k}\left(yz_k\right)\right]. 
\end{align*}
Since $Y_{k+1}>Y_k$ , we select $z_k =\frac{\mathbb{E}[Y_{k}]}{\sum_{i=1}^{K}\mathbb{E}[Y_i]}= \frac{2k}{K(K+1)}$ to get a tighter lower bound. Then substituting the above $z_k$ into the above result in step $(d)$ yields the upper bound on $\mathbb{P}[S_K\geq y]$ in \eqref{Eqn:LowBoundSK}. 

\subsection{Proof of Lemma \ref{Lem:NoUplinkColProb}} \label{App:ProofNoUplinkColProb}
Let $M_k$ denote the number of the users associating with the $k$th AP in the virtual cell. The probability of no collisions happening in the cell of the $k$th AP is $\delta (1-\delta)^{M_k-1}$ if the probability of a user selecting any one of the RRUs for each AP is $\delta$. Suppose the radio resource (available bandwidth) of each AP can be divided into $R$ radio resource units so that we have $\delta=\frac{1}{R}$. Thus, the probability that an AP in the virtual cell does not have collisions, denoted by  $\rho^{ul}$, can be written as
\begin{align*}
\rho^{ul} &=\sum_{r=1}^R \delta \, \mathbb{E}\left[(1-\delta)^{M_k-1}\right]=\sum_{r=1}^R \frac{1}{R} \, \mathbb{E}\left[(1-\delta)^{M_k-1}\right]\\
&=\mathbb{E}\left[(1-\delta)^{M_k-1}\right],
\end{align*} 
where $\mathbb{E}[(1-\delta)^{M_k-1}]=\sum_{m=1}^{2}\mathbb{P}[V_k\in\Phi_m]\mathbb{E}[(1-\delta)^{M_k-1}|V_k\in\Phi_m]$ and we thus have
\begin{align*}
\rho^{ul} &=\sum_{m=1}^{2}\vartheta_m\mathbb{E}\left[(1-\delta)^{N_m-1}\right]\\
&=\sum_{m=1}^{2}\vartheta_m\sum_{n=1}^{\infty}p_{m,n}(1-\delta)^{n-1},
\end{align*}
where $\vartheta_m=\mathbb{P}[V_k\in\Phi_m]$ and $N_m$ is the number of the users associating with a tier-$m$ AP.  Moreover, we know that the probability that the nearest AP to the user located at the origin is from the $m$th tier is $\mathbb{P}[\|A_{m,*}\|^{-\alpha}\geq \|A_{i,*}\|^{-\alpha}]=\mathbb{P}[\|A_{m,*}\|^2\leq \|A_{i,*}\|^2]$ for $m\neq i$ in which $A_{i,*}$ is the nearest point in $\Phi_i$ to the user. Since $c^{-1}\|A_{m,*}\|^2\sim\exp(c\pi\lambda_m)$ for any $c>0$, we thus have $\vartheta_m =\mathbb{P}\left[\|A_{m,*}\|^2\leq \|A_{k,*}\|^2\right]=\frac{\lambda_m}{\sum_{i=1}^{m}\lambda_i}$.
Substituting the above result of $\vartheta_m$ into the above expression of $\rho^{ul}$ yields the result in \eqref{Eqn:NonCollProbEachAP}. In addition, the uplink non-collision reliability of the user is the probability that there is at least one non-collision AP in the virtual and it thus can be expressed as $\rho^{ul}_K=1-(1-\rho^{ul})^K$, which is equal to the result in \eqref{Eqn:NonColProb} by substituting the result of $\rho^{ul}$ in \eqref{Eqn:NonCollProbEachAP} into $\rho^{ul}_K$.  

\subsection{Proof of Theorem \ref{Thm:UplinkCommRel}}\label{App:ProofUplinkCommRel}
First consider the scenario in which all $K$ APs in the virtual cell do not collaborate and the transmit power $q$ of users is equally allocated to the $K$ APs, i.e., $q_i=q_k=q/K$ for all $i\in\mathbb{N}_+$ and $k\in\{1,\ldots,K\}$. If all the non-collision uplink SIRs in \eqref{Eqn:UplinkCommRel} are independent, we have
\begin{align*}
\eta^{ul}_K =&1-\mathbb{P}\left[\max_{k\in\{1,\ldots,K\}}\{\gamma^{ul}_k\mathds{1}(V_k\in \mathcal{V}^{nc}_K)\}\leq \theta\right]\\
\approx & 1-\prod_{k=1}^K \big(\mathbb{P}\left[\gamma^{ul}_k\leq \theta\right]\mathbb{P}[\mathds{1}(V_k\in \mathcal{V}^{nc}_K)=1]\\
&+1-\mathbb{P}[\mathds{1}(V_k\in \mathcal{V}^{nc}_K)=1]\big)\\
\stackrel{(a)}{=}& 1-\prod_{k=1}^K \left\{1-\rho^{ul}\mathbb{P}\left[\gamma^{ul}_k\geq\theta\right]\right\},
\end{align*}
where $(a)$ follows from the result of $\mathbb{P}[\mathds{1}(V_k\in \mathcal{V}^{nc}_K)=1]=\sum_{m=1}^{2}\vartheta_m\sum_{n=1}^{\infty}p_{m,n}(1-\delta)^{n-1}=\rho^{ul}$. Whereas $\mathbb{P}[\gamma^{ul}_k\geq\theta]$ can be derived as shown in the following:
\begin{align*}
\mathbb{P}[\gamma^{ul}_k\geq\theta] &= \mathbb{P}\left[\frac{h_k\|V_k\|^{-\alpha}}{\sum_{j:U_j\in\mathcal{U}_{a}}h_j\|V_k-U_j\|^{-\alpha}}\geq\theta\right]\\
&\stackrel{(b)}{=}\mathbb{E}\left[\exp\left(-\theta\|V_k\|^{\alpha}\sum_{j:U_j\in\mathcal{U}_a}\frac{h_j}{\|U_j\|^{\alpha}}\right)\right]\\
&\stackrel{(c)}{=}\mathbb{E}_{\|V_k\|^2}\left[\exp\left(-\frac{\pi \theta^{\frac{2}{\alpha}} \|V_k\|^2 \mu_a}{\mathrm{sinc}(2/\alpha)}\right)\right]\\
&\stackrel{(d)}{=}\left(1+\frac{ \theta^{\frac{2}{\alpha}}\mu_a}{\mathrm{sinc}(2/\alpha)\widetilde{\lambda}}\right)^{-k},
\end{align*}
where $(b)$ follows from $h_k\sim\exp(1)$ and the Slivnyak theorem saying that the statistical property of a homogeneous PPP evaluated at $V_k$ is the same as that evaluated at the origin (or any point in the network) \cite{DSWKJM13,FBBBL10}, $(c)$ is obtained by first applying the PGFL of a homogeneous PPP to $\mathcal{U}_a$ that is a homogeneous PPP of density $\mu_a=\delta(1-p_0)\sum_{m=1}^{2}\lambda_m=\delta(1-p_0)\widetilde{\lambda}$, and $(d)$ is due to $\|V_k\|^2\sim \text{Gamma}(k,\pi\widetilde{\lambda})$. Then substituting the result of $\mathbb{P}[\gamma^{ul}_k\geq\theta]$ into the result of $\eta^{ul}_K$ found in $(a)$ leads to the result in \eqref{Eqn:UplinkCommRelRel}. Next, we consider the case in which a user can associate with all APs in the network, i.e., $K$ goes to infinity in this case, and we would like to find $\rho^{ul}_{\infty} \defn \lim_{K\rightarrow\infty} \rho^{ul}_K$. We use the above results in $(a)$ and $(c)$ to approximately express $\rho^{ul}_{\infty}$ as
\begin{align*}
\rho^{ul}_{\infty} &\approx 1-\mathbb{E}\left\{\prod_{V_k\in\Phi} \left(1-\rho^{ul}\mathbb{P}\left[\gamma^{ul}_k\geq\theta|V_k\right]\right)\right\}\\
&=1-\exp\left\{-\pi\rho^{ul}\widetilde{\lambda} \int_{0}^{\infty} e^{-\frac{\pi \theta^{\frac{2}{\alpha}} r \mu_a}{\mathrm{sinc}(2/\alpha)}} \dif r \right\},
\end{align*}
which is equal to the result in \eqref{Eqn:UplinkCommRelManyAPs} by carrying out the integral in the last equality for $\mu_a=\delta\widetilde{\lambda}$. 

Next we would like to find the bounds on $\eta^{ul}_K$ when all the non-collision APs in the virtual cell are able to collaborate. The upper bound on $\eta^{ul}_K$ in \eqref{Eqn:UplinkCommRelCoop} can be thereupon derived as follows:
\begin{align*}
\eta^{ul}_K =& \mathbb{P}\left[\frac{\sum_{k:V_k\in\mathcal{V}_K}h_kq_k\|V_k\|^{-\alpha}\mathds{1}(V_k\in\mathcal{V}^{nc}_K)}{\sum_{j:U_j\in\mathcal{U}_{a}\setminus\mathcal{V}_K}h_jq_j\|V_k-U_j\|^{-\alpha}}\geq\theta \right]\\
\stackrel{(e)}{=}&  \rho^{ul}\mathbb{P}\left[\sum_{k:V_k\in\mathcal{V}_K}\frac{h_k}{\|V_k\|^{\alpha}}\geq \theta \sum_{j:U_j\in\mathcal{U}_{a}}\frac{h_j}{\|U_j\|^{\alpha}}\right]\\
\stackrel{(f)}{\leq} & \rho^{ul}\bigg\{1-\prod_{k=1}^{K}\bigg(1-\mathbb{E}_{Y_k}\bigg[\exp\bigg(- \frac{\pi\mu_aY_k}{ \text{sinc}(2/\alpha)}\\
&\times\left(\frac{2k\theta}{K(K+1)}\right)^{\frac{2}{\alpha}}\bigg)\bigg]\bigg)\bigg\},
\end{align*} 
where $(e)$ follows from $\mathbb{P}[\mathds{1}(V_k\in\mathcal{V}^{nc}_K)]=\rho^{ul}$ and $q_j=q/K$ for all $j$, and $(f)$ follows from the result in \eqref{Eqn:LowBoundSK} for $w_m=1$. Thence, using $\mathbb{E}_{Y_k}[\exp(-sY_k)]=(1+s/\pi\widetilde{\lambda})^{-k}$ for $s>0$ and substituting $\mu_a=\delta(1-p_0)\widetilde{\lambda}$ into the above result of $\eta^{ul}_K$ yield the result in \eqref{Eqn:UplinkRelCop}.  

\subsection{Proof of Theorem \ref{Thm:DownlinkReliability}}\label{App:ProofDownlinkReliability}
(i) Since the intra-virtual-cell interference exits in the virtual cell, we know that $\gamma^{dl}_k$ in \eqref{Eqn:Downlink-kthSIR} can be equivalently expressed as $\gamma^{dl}_k \stackrel{d}{=} \frac{H_k\|\widetilde{V}_k\|^{-\alpha}}{\sum_{i:\widetilde{V}_i\in\widetilde{\mathcal{V}}_K\setminus \widetilde{V}_k}H_i\|\widetilde{V}_i\|^{-\alpha}+I^{dl}_K}\leq \frac{H_k\|\widetilde{V}_k\|^{-\alpha}}{\widetilde{I}^{dl}_k}$
where $\widetilde{I}^{dl}_k\defn \sum_{j:\widetilde{V}_j\in\widetilde{\Phi}\setminus\widetilde{\mathcal{V}}_k}\frac{O_jH_j}{\|\widetilde{V}_j\|^{\alpha}}$, $\widetilde{\Phi}$, $\widetilde{V}_k$, and $\widetilde{V}_k$ are already defined in Appendix \ref{App:ProofLapTransSK} and $O_j$ is the Bernoulli random variable associated with $\widetilde{V}_j$ and it is unity if $\widetilde{V}_j$ is not void and zero otherwise. Note that  the inequality comes from the fact that $\widetilde{I}^{dl}_k$ is smaller than $\sum_{i:\widetilde{V}_i\in\widetilde{\mathcal{V}}_K\setminus \widetilde{V}_k}H_i\|\widetilde{V}_i\|^{-\alpha}+I^{dl}_K$ because it does not contain the interference from the first $k$ APs. Thus, $\eta^{dl}_K=1-\mathbb{E}_{\mathcal{V}_K}\left\{\prod_{k=1}^{K}\mathbb{P}\left[\gamma^{dl}_k\leq\theta|\mathcal{V}_{K}\right] \right\}$ is upper bounded by
\begin{align*}
\eta^{dl}_K & \leq 1-\mathbb{E}_{\widetilde{\mathcal{V}}_k}\left\{\prod_{k=1}^K \mathbb{P}\left[ \frac{H_k\|\widetilde{V}_k\|^{-\alpha}}{\widetilde{I}^{dl}_k}\leq \theta\bigg| \widetilde{\mathcal{V}}_k\right]\right\}\\
&\stackrel{(a)}{=} 1- \prod_{k=1}^{K}\left\{1- \mathbb{E}\left[e^{-\theta \|\widetilde{V}_k\|^{\alpha} \sum_{j:\widetilde{V}_j\in\widetilde{\Phi}}\frac{O_jH_j}{\|\widetilde{V}_j\|^{\alpha}}}\right]\right\},
\end{align*}
where $(a)$ is obtained by the assumption that all $\gamma^{dl}_k$'s are independent and $H_k\sim\exp(1)$. According to the result in \eqref{Eqn:WeighedSK} and the density of the APs in $\widetilde{\Phi}$ that uses the same RRU is $\delta\widetilde{\lambda}$, the expectation in step $(a)$ can be simplified as follows:
\begin{align*}
& \mathbb{E}\left\{\exp\left[-\theta \sum_{j:\widetilde{V}_j\in\widetilde{\Phi}}O_jH_j\left(\frac{\|\widetilde{V}_j\|^2}{\|\widetilde{V}_k\|^2}\right)^{-\frac{\alpha}{2}} \right] \right\}\\
&=\left\{1+\ell\left(\theta,\frac{2}{\alpha}\right)\delta\mathbb{E}\left[O^{\frac{2}{\alpha}}\right]\right\}^{-k}.
\end{align*}
Thus, we have the upper bound $\eta^{dl}_K \leq 1-\prod_{k=1}^{K} \left\{1-\left(1+\delta\ell\left(\theta,\frac{2}{\alpha}\right)\mathbb{E}\left[O^{\frac{2}{\alpha}}\right]\right)^{-k}\right\}$,
which yields the result in \eqref{Eqn:UppBoundDLRel} because   $\mathbb{E}[O^{2/\alpha}]=\mathbb{P}[O=1]=\sum_{m=1}^{2}\vartheta_m(1-p_{0,m})$.

(ii) Now consider the case of $K\rightarrow\infty$. In this case, all APs in the network are associated with the typical user and $\gamma_k$ in \eqref{Eqn:Downlink-kthSIR} can be rewritten as $\gamma^{dl}_k=\frac{H_kQ_k\|V_k\|^{-\alpha}}{\sum_{i:V_i\in\mathcal{V}_{\infty}\setminus V_k}H_iQ_i\|V_i\|^{-\alpha}}\stackrel{d}{=} \frac{H_k\|\widetilde{V}_k\|^{-\alpha}}{\widetilde{I}^{dl}_k}$
where $\widetilde{I}^{dl}_k\defn \sum_{i:\widetilde{V}_i\in\widetilde{\mathcal{V}}_{\infty}\setminus \widetilde{V}_k}H_i\|\widetilde{V}_i\|^{-\alpha}$. For a given $\|\widetilde{V}_k\|^2=r$ and $H_k\sim\text{exp}(1)$, we have $\mathbb{P}[\gamma^{dl}_k\geq\theta]=\mathbb{P}\left[H\geq \theta r^{\frac{\alpha}{2}}\widetilde{I}^{dl}_k\right]=\exp\left(-\frac{\pi\widetilde{\lambda}\theta^{\frac{2}{\alpha}}r}{\text{sinc}(2/\alpha)}\right)$.
Thus, $\eta^{dl}_{\infty}$ can be expressed and derived as follows:
\begin{align*}
\eta^{dl}_{\infty} &= 1-\mathbb{P}\left[ \max_{k:\widetilde{V}_k\in\widetilde{\mathcal{V}}_{\infty}}\{\gamma^{dl}_k\} \leq\theta\right]\\ 
&\stackrel{(b)}{=} 1-\exp\left[-\pi\widetilde{\lambda}\int_{0}^{\infty}\exp\left(-\frac{\pi\delta\widetilde{\lambda}\theta^{\frac{2}{\alpha}}r}{\text{sinc}(2/\alpha)}\right) \dif r \right],
\end{align*}
where $(b)$ follows by assuming all $\gamma_k$'s are independent and then finding the PGFL of $\widetilde{\mathcal{V}}_{\infty}$. Carrying out the integral in the step of $(b)$ leads to the result in \eqref{Eqn:BoundDownConnRel}.

(iii) When all the $K$ APs in the virtual cell can collaborate, $\eta^{dl}_K$ in \eqref{Eqn:RedDownlinkSIRCoop} can be rewritten as
\begin{align*}
\eta^{dl}_K =& \mathbb{P}\left[\frac{\sum_{k=1}^{K}H_k\|\widetilde{V}_k\|^{-\alpha}}{\sum_{i:\widetilde{V}_i\in\widetilde{\Phi}\setminus\widetilde{\mathcal{V}}_K}O_iH_i\|\widetilde{V}_i\|^{-\alpha}}\geq\theta\right]\\
\stackrel{(c)}{\leq}& 1- \mathbb{P}\left[K\max_{k}\{H_k\}\leq\theta\|\widetilde{V}_1\|^{\alpha}\sum_{i:\widetilde{V}_i\in\widetilde{\Phi}\setminus\widetilde{\mathcal{V}}_K}
\frac{O_iH_i}{\|\widetilde{V}_i\|^{\alpha}}\right]\\
\stackrel{(d)}{=}& 1-   \left\{1-\mathbb{E}\left[\exp\left(-\frac{\theta\|\widetilde{V}_K\|^{\alpha}}{K^{\frac{\alpha}{2+1}}}\sum_{i:\widetilde{V}_i\in\widetilde{\Phi}\setminus\widetilde{\mathcal{V}}_K}\frac{O_iH_i}{\|\widetilde{V}_i\|^{\alpha}}\right)\right]\right\}^K\\
\stackrel{(e)}{=}&1-\bigg\{1-\left[1+\delta\ell\left(\frac{\theta}{K^{\frac{\alpha}{2}+1}},\frac{2}{\alpha}\right)\sum_{m=1}^{2}\vartheta_m(1-p_{m,0})\right]^K \\
&\bigg\}^{-K},
\end{align*}  
where $(c)$ is because $(\min_{k}H_k)\|\widetilde{V}_K\|^{-\alpha}<H_k\|\widetilde{V}_k\|^{-\alpha}$ for all $k<K$, $(d)$ is due to $\mathbb{P}[\min_{k}\{H_k\}\leq x]=1-\exp(-Kx)$ and replacing set $\widetilde{\Phi}\setminus \widetilde{\mathcal{V}}_K$ with set $\widetilde{\Phi}\setminus \widetilde{\mathcal{V}}_1$, and $(e)$ follows from the result in \eqref{Eqn:WeighedSK} for setting $K$ as unity and $\theta$ as $\frac{\theta}{K}$. Hence, \eqref{Eqn:RelDownLinkCoop} is acquired and the proof is complete.
\bibliographystyle{IEEEtran}
\bibliography{IEEEabrv,Ref_UltraReliableHetNet}

\end{document}